\documentclass[prb,twocolumn]{revtex4}
\usepackage{amssymb}
\usepackage{graphicx}

\def\v#1{\textbf{\emph{#1}}}
\def\c#1{\mathbb{#1}}
\def\t#1{\widetilde{#1}}

\def\skiip#1{}

\begin{document}
\title{
Tensor-Entanglement-Filtering Renormalization Approach\\
and Symmetry Protected Topological Order
}
\author{Zheng-Cheng Gu}
\author{Xiao-Gang Wen}
\affiliation{
Department of Physics, Massachusetts Institute of
Technology, Cambridge, Massachusetts 02139, USA
}
\date{{Sept., 2008}}

\begin{abstract}
We study the renormalization group flow of the Lagrangian for
statistical and quantum systems by representing  their path integral
in terms of a tensor network.  Using a tensor-entanglement-filtering
renormalization (TEFR) approach that removes local entanglement and
coarse grain the lattice, we show that the resulting renormalization
flow of the tensors in the tensor network has a nice fixed-point
structure.  The isolated fixed-point tensors $T_\text{inv}$ plus the
symmetry group $G_\text{sym}$ of the tensors (\ie the symmetry group
of the Lagrangian) characterize various phases of the system.  Such a
characterization can describe both the symmetry breaking phases and
topological phases, as illustrated by 2D statistical Ising model, 2D
statistical loop gas model, and 1+1D quantum spin-1/2 and spin-1
models.  In particular, using such a $(G_\text{sym}, T_\text{inv}) $
characterization, we show that the Haldane phase for a spin-1 chain is
a phase protected by the time-reversal, parity, and translation
symmetries.  Thus the Haldane phase is a symmetry protected
topological phase.  The $(G_\text{sym}, T_\text{inv})$
characterization is more general than the characterizations based on
the boundary spins and string order parameters.  The tensor
renormalization approach also allows us to study continuous phase
transitions between symmetry breaking phases and/or topological
phases.  The scaling dimensions and the central charges for the
critical points that describe those continuous phase transitions can
be calculated from the fixed-point tensors at those critical points.
\end{abstract}

\maketitle


\tableofcontents


\section{Introduction}

For a long time, people believe that all phases and all continuous
phase transitions are described by symmetry breaking.\cite{L3726}
Using the associated concepts, such as order parameter, long range
correlation and group theory, a comprehensive theory for symmetry
breaking phases and their transitions is developed.  However, nature's
richness never stops to surprise us.  In a study of chiral spin liquid
in high temperature superconductor,\cite{KL8795,WWZcsp} we realized
that there exist a large class of phases beyond symmetry
breaking.\cite{Wtop} Since the new phases can be characterized by
topology-dependent topologically stable ground state degeneracies, the
new type of orders in those phases are named topological
orders.\cite{Wrig}.  Later, it was realized that fractional quantum
Hall states\cite{TSG8259} are actually topologically ordered
states.\cite{WNtop}

Recently, it was shown that topological order is actually the
pattern of long range entanglement in the quantum ground
state.\cite{Wqogen,KP0604,LWtopent} The possible patterns of long
range entanglement are extremely rich,\cite{FNS0428,LWstrnet,NSS0789}
which indicate the richness of new states of matter beyond the
symmetry breaking paradigm.  Such rich new states of matter deserve a
comprehensive theory.\cite{Wtoprev,Wen04}

But how to develop a comprehensive theory of topological order? First
we need a quantitative and concrete description of topological order.
The topological ground state degeneracy\cite{Wtop,WNtop} is not enough
since it does not completely characterize topological order.  However,
the non-Abelian Berry's phases from the degenerate ground states and
the resulting representation of modular group provide a more complete
characterization of topological order.\cite{Wrig} It may even
completely characterize topological order.  There are other ways to
(partially)  characterize topological order, such as through the
fractional statistics of the quasiparticles\cite{ASW8422,K062} or
through the string operators that satisfy the
``zero-law''.\cite{HWcnt} Those ``zero-law''  string operators appear
to be related to another characterization of topological order through
a gauge-like-sysmetry defined on a lower dimensional subspace of the
space.\cite{NO0616,CN0801} 

But so far, we have not been able to use those characterizations to
develop a theory of topological order.  So the key to develop a
comprehensive theory of topological order is to find a quantitative
and easy-to-use description of topological order.  For fractional
quantum Hall states, the pattern of zeros appears to provide a
quantitative and easy-to-use description of topological
order.\cite{WWsymm,WWsymmqp,BW0889} For more general systems, we may
use the renormalization group (RG) approach to find a quantitative and
easy-to-use description.  We will see that the new description based
on the RG approach naturally leads to a theory of topological order
and phase transitions between topologically ordered states.

Renormalization group (RG) approach\cite{K6663,W7174} is a powerful
method to study phases and phase transitions for both quantum and
statistical systems.  For example, for a quantum system, one starts
with its Lagrangian and path integral that describe the system at a
certain cut-off scale. Then one integrates out the short distance
fluctuations in the path integral to obtain a renormalized effective
Lagrangian at a longer effective cut-off scale.  Repeating such a
procedure, the renormalized effective Lagrangian will flow to an
isolated fixed-point Lagrangian that represents a phase of the system.
As we adjust coupling constants in the original Lagrangian, we may
cause the renormalization flow to switch from one fixed point to
another fixed point, which will represent a phase transition.  Since
the fixed-point Lagrangian is unique for each phase and the phase
transition is associated with a change in fixed-point  Lagrangian, it
is natural to use the fixed-point Lagrangian to describe various phases.

But can we use the fixed-point Lagrangian to quantitatively describe
topological order? At first sight, the answer appears to be no.  When a
quantum spin system (or a qbit system) has a topologically ordered
ground state, its low energy effective Lagrangian (\ie the
fixed-point Lagrangian) can be a $Z_2$ gauge
theory,\cite{RS9173,Wsrvb,MS0181,Wqoslpub,K032} a $U(1)$ gauge
theory,\cite{MS0204,Wqoslpub,Wlight,MS0312} a $SU(2)$ gauge
theory,\cite{Wqoslpub} a QED theory (with $U(1)$ gauge field and
massless fermion field),\cite{Wqoslpub,Wqoem,LWqed} a QCD theory (with
non-Abelian gauge fields and massless fermion
fields),\cite{DFM8826,AZH8845,Wqoslpub,Wqoem} a Chern-Simons
theory,\cite{WWZcsp,Wnab} \etc.  In the standard implication of RG
approach (within the frame work of field theory), a bosonic Lagrangian
that describes a quantum spin system cannot flow to all those very
different possible fixed-point Lagrangian with very different fields
(such as gauge fields and anticommuting fermionic fields).

Since lattice spin systems do have those very different topologically
ordered phases, we see that the standard implementation of the RG
approach is incomplete and needed to be generalized to cover those
existing (and exciting) renormalization flows that end at very
different fixed-point Lagrangian. In this paper, we will describe one
such generalization -- tensor network renormalization approach -- that
should allow us to study any phases (both symmetry breaking and
topological phases) of any systems (with a finite cut-off).  The
resulting fixed-point Lagrangian from the tensor network
renormalization approach give us a quantitative description of
topological orders (and the symmetry breaking orders).

The tensor network renormalization approach is based on an observation
that the space-time path integral of a quantum spin system or the
partition function of a statistical system on lattice can be
represented by a tensor trace over a tensor network after we
discretize the time and the fields:
\begin{align}
 \int D \vphi(\v x,t) \e^{-\int \cL}
=\sum_{\{ \phi_{\v i} \} }  \prod T_{
\phi_{\v i}
\phi_{\v j}
\phi_{\v k}
\phi_{\v l}}\equiv \text{tTr} \otimes_{\v i} T
\end{align}
where $\v i$ label the discretized space-time points and $\phi_{\v i}$
is the discretized field (for detail, see \eq{tTrsq}).  The different
choices of tensor $T_{ \phi_{\v i} \phi_{\v j} \phi_{\v k} \phi_{\v
l}}$ in the tensor network correspond to different
Lagrangians/partition-functions.  We like to point out that in
addition to use it to describe path integral or partition function,
tensor network can also be used to describe many-body wave
functions.\cite{VC0466}  Based on such a connection, many variational
approaches were
developed.\cite{VC0466,MVC0506,V0705,JOV0802,JWX0803,PEV0880,GLWtergV}

Since every physical system is described by path integral, so if we
can calculate the tensor trace, we can calculate anything.
Unfortunately, calculating the tensor-trace $\text{tTr}$ is an NP hard
problem in 1+1D and higher dimensions.\cite{SWV0706}  In \Ref{LN0701},
Levin and Nave introduced a tensor renormalization group procedure
where a coarse graining of the tensor network is performed to produce
a new tensor on the coarse grained network.  This leads to an
efficient and accurate calculation path integral (or partition
function) and other measurable quantities.  Such a method opens a new
direction in numerical calculation of many-body systems.

The above tensor renormalization group procedure also gives rise to a
renormalization flow of the tensor.  However, the fixed-point tensor
obtained from such a flow is not isolated: even starting with different
Lagrangian (\ie the tensors) that describe the same phase, the
renormalization flow of the those Lagrangian will end up at different
fixed-point tensors.  This means that the renormalization procedure
does not filter out (or integrate out) all the short distance
fluctuations so that the fixed-point tensors still contain short
distance non-universal information.  In this paper, we will introduce
an improved tensor renormalization group procedure by adding a
filtering operation that remove the local entanglements.  We show that
this additional entanglement filtering procedure fixes the above
problem, at least for the cases studied in this paper.

To test the new tensor network renormalization method, 
we study the fixed-point tensors that
correspond to the two phases of 2D statistical Ising model which has a
$Z_2$ spin flip symmetry.  We find that the fixed-point tensor for
the disordered phase is a trivial tensor $T^\text{TRI}$ where the
dimensions of all the tensor indices are equal to one (we will refer
such a trivial tensor as a dimension-one tensor which is represented
by a number).  While the fixed-point tensor for the symmetry breaking
(spin ordered) phase is a dimension-two tensor $T^{Z_2}$ which is a
direct sum of two dimension-one tensors: $T^{Z_2}=T^\text{TRI}\oplus
T^\text{TRI}$.  We believe that this is a general feature for symmetry
breaking phase: a symmetry breaking phase
with $n$ degenerate states is represented by fixed-point tensors that
is a direct sum of $n$ dimension-one trivial tensors.

Since the $Z_2$ spin flip symmetry is crucial for the existence and
the distinction of the disordered phase and the spin ordered phase, we
should use a pair, the symmetry group $G_\text{sym}=Z_2$ and the
fixed-point tensors $T_\text{inv}=T^\text{TRI},\ T^{Z_2}$, to
characterize the two phases. So the disordered phase is characterized
by $(Z_2, T^\text{TRI})$ and the spin ordered phase is characterized by
$(Z_2, T^{Z_2})$. It turns out that the characterization based on the
symmetry group and the fixed-point tensor, $(G_\text{sym},T_\text{inv})$,
is very general. It can describe both symmetry breaking phases and
topological phases. This is one of the main results of this paper.

Even at the critical point between the symmetry breaking phases and/or
topological phases, the tensor renormalization approach produces fixed-point
tensors (which now have an infinite dimension).  From those fixed-point
tensors, or more precisely, from a finite-dimension approximation of those
fixed-point tensors, we can calculate the critical properties of the critical
points with an accuracy of 0.1\%.  Using such an method, we confirm that the
phase transition between the disordered and ordered phases of the Ising model
is a central charge $c=1/2$ critical point.

We also study a 2D statistical loop gas model.  We find that for the
small-loop phase, the fixed-point tensor is the dimension-one tensor
$T^\text{TRI}$, while for the large loop phase, the fixed-point tensor
$T^\text{LL}$ is a tensor with dimension 2 (see \eq{Tll}).  However,
unlike the symmetry breaking phase, the fixed-point tensor for the
large loop phase is \emph{not} a direct sum of two dimension-one
tensors: $T^\text{LL}\neq T^\text{TRI}\oplus T^\text{TRI}$.  This is
consistent with the fact that, as a \emph{statistical} system, the
small-loop phase and the large loop phase has the same symmetry and
the phase transition is not a symmetry breaking transition.  The
example demonstrates that fixed-point tensors can be used to study phases
and phase transitions beyond symmetry breaking.  The phase transition
between the two phases is again a central charge $c=1/2$ critical
point.

Next, we study a spin-1/2 model $H= \sum_i (-\si^x_i\si^x_{i+1} -J
\si^z_i\si^z_{i+1} )$, with $Z^x_2\times Z^z_2$ symmetry generated by
$\imth\si^x$ and $\imth\si^z$.  The spin-1/2 model has two different $Z_2$
symmetry breaking phases described by, in terms of the new notation,
$(Z^x_2\times Z^z_2, T^{Z_2})$ and $(Z^x_2\times Z^z_2, T^\text{LL})$.
The $(Z^x_2\times Z^z_2, T^{Z_2})$ phase breaks the $Z^x_2$ symmetry
while the $(Z^x_2\times Z^z_2, T^\text{LL})$ phase breaks the $Z^z_2$
symmetry.  The phase transition between the two $Z_2$ symmetry
breaking phases is a continuous phase transition that is beyond the
Landau symmetry breaking paradigm.  We find that the tensor network
renormalization approach can
describe such a continuous phase transition.  The phase transition
between the two phases is a central charge $c=1$ critical point.

Last we study the $T=0$ phases of quantum spin-1 model in 1+1D $
H=-\sum_{i} [ \v S_{i}\cdot \v S_{i+1} + U (S^z_{i})^2 + B S^x_i]$,
and found that the model has four different phases for different
$(U,B)$: a trivial phase, two $Z_2$ symmetry breaking phases, and the
Haldane phase.\cite{H8364,H8353,AKL8799} We find that the Haldane
phase is described by a dimension-four
fixed-point tensor $T^\text{Haldane}$ (see \eq{THaldane}).
$T^\text{Haldane}$ is not a direct sum of dimension-one tensors.  From
the fixed-point tensor $T^\text{Haldane}$, we confirm that the Haldane
phase has a finite energy gap and the ground state is not degenerate.
Thus, despite the fixed-point tensor to be a dimension-four tensor,
the Haldane phase, like the trivial phase, does not break any symmetry
(as expected).

The TEFR approach allows us to study the stability of the Haldane
phase (\ie the distinction between the Haldane phase and the trivial
phase) under a very general setting.  Our calculation shows that the
fixed-point tensor $T^\text{Haldane}$, and hence, the Haldane phase is
robust against any perturbations that have time-reversal, parity, and
translation symmetries.\footnote{When we say the Haldane phase is
robust against certain perturbations, we mean that, even in the
presence of those perturbations, the Haldane phase is still separated
from the trivial $S^z=0$ phase by phase transitions.}  So the Haldane
phase is a symmetry protected topological phase characterized by the
pair $(G_{TPT}, T^\text{Haldane})$ where $G_{TPT}$ is the symmetry
group generated by time-reversal, parity, and translation symmetries.

People have been using the degenerate spin-1/2 degrees of freedom at
the ends of spin-1 chain,\cite{HKA9081,GGL9114,N9455} and string order
parameter\cite{NR8909,KT9204} to characterize Haldane phase.  Both
degenerate boundary spin-1/2 degrees of freedom and the string order
parameter can be destroyed by perturbations that break spin rotation
symmetry but have time-reversal, parity, and translation symmetries.  So
if we used the degenerate spin-1/2 degrees of freedom and/or string
order parameter to characterize the Haldane phase, we would
incorrectly conclude that the Haldane phase is unstable against those
perturbations.  Since the Haldane phase is indeed stable against those
perturbations, we conclude that the $(G_{TPT}, T^\text{Haldane})$
characterization is more general.  We also find that an arbitrary
small perturbation that break parity symmetry will destabilize the
Haldane phase.  This confirms a previously known result.\cite{BTG0819}

In the first a few sections of this paper, we will review and
introduce several tensor network renormalization approaches.  Then we
will discuss new physical picture about phases and phase transitions
emerged from the tensor network renormalization approach through
several simple examples.

\begin{figure}
\begin{center}
\includegraphics[scale=0.6]{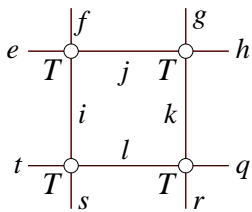}
\end{center}
\caption{
A graphic representation of a tensor-network on a square lattice.  A
vertex represent a tensor $T_{ijfe}$ and the legs of a vertex carries
the indices $i$, $j$, \etc. Each link carries the same index.  The
indices on the internal links are summed over, which defines the
tensor-trace. }
\label{tnsq}
\end{figure}

\section{A review of a tensor network renormalization method}

In the following review, we will concentrate on 1+1D quantum systems.
However, our discussion can be generalized to 2D statistical systems
and the systems in higher dimensions.

To describe the tensor renormalization approach introduced
in \Ref{LN0701}, we first note that, after we discretize
the space-time, the partition function represented by a space-time
path integral $\Tr \e^{-\bt H}=\int \e^{-\int \cL}$ (in imaginary
time) can be written as a tensor-trace over a tensor-network (see Fig.
\ref{tnsq} for an example on 2D square lattice):
\begin{equation}
\label{tTrsq}
\Tr \e^{-\bt H}
=\sum_{ijkl\cdots}
T_{jfei}T_{hgjk}T_{qklr}T_{lits}\cdots
=\text{tTr} \otimes_{\v i} T.
\end{equation}
where the indices of the tensor run from $1$ to $D$ (in other words
the tensor $T$ is a rank-four dimension-$D$ tensor).  Choosing
different tensors $T$ corresponds to choosing different Lagrangian
$\cL$.  We see that calculating the tensor trace allows us to
obtain properties of any physical systems.

\begin{figure}
\begin{center}
\includegraphics[scale=0.5]
{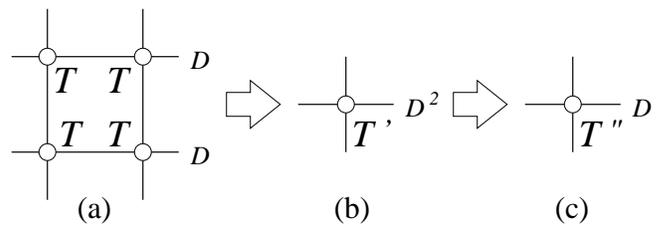}
\end{center}
\caption{
The tensor $T$ in the tensor network (a) has a dimension $D$.  After
combine the two legs on each side into a single leg, the four linked
tensors in (a) can be viewed as a single tensor $ T'$ in (b) with
dimension $D^2$. $T'$ can be approximately reduced to a ``smaller''
tensor $T^{\prime\prime}$ in (c) with dimension $D$ and satisfies $
\text{tTr} [ T^{\prime}\otimes  T^{\prime} \cdots] \approx \text{tTr}
[ T^{\prime\prime}\otimes  T^{\prime\prime} \cdots] $.
}
\label{RG}
\end{figure}

Unfortunately, calculating the tensor-trace $\text{tTr}$ is an NP hard
problem in 1+1D and higher dimensions.\cite{SWV0706}  To solve this
problem, \Ref{LN0701} introduced an approximate real-space
renormalization group approach which accelerate the calculation
exponentially.  The basic idea is quite simple and is illustrated in
Fig. \ref{RG}.  After finding the reduced tensor $T^{\prime\prime}$,
we can express $\text{tTr}[T\otimes T\cdots] \approx
\text{tTr}[T^{\prime\prime}\otimes T^{\prime\prime}\cdots ]$ where the
second tensor-trace only contains a quarter of the tensors in the
first tensor-trace.  We may repeat the procedure until there are only
a few tensors in the tensor-trace. This allows us to reduce the
exponential long calculation to a polynomial long calculation.

\begin{figure}
\begin{center}
\includegraphics[width=3.5in]
{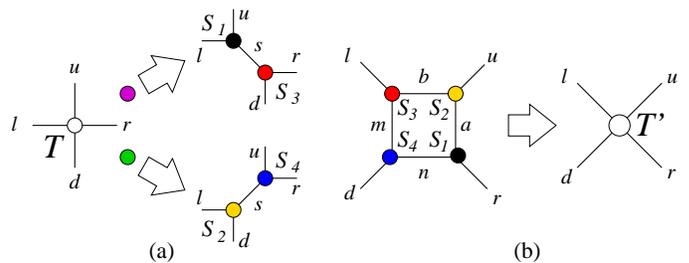}
\end{center}
\caption{(Color online) (a) We represent the original rank-four tensor by two
rank-three tensors, which is an \emph{approximate} decomposition.
(b) Summing over the indices around the square produces a single tensor
$T'$.  This step is \emph{exact}.
} \label{tsrd}
\end{figure}

\begin{figure}
\begin{center}
\includegraphics[width=3.5in]
{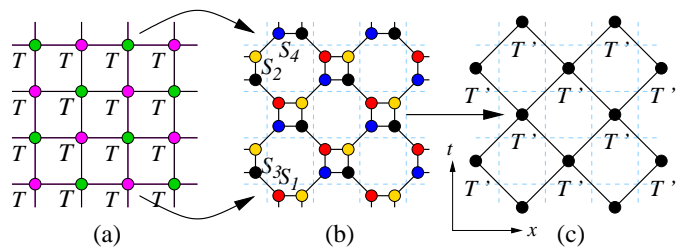}
\end{center}
\caption{(Color online) RG transformation of tensor-network produces a
coarse grained tensor-network.
}
\label{CGsq}
\end{figure}

The actual implementation of the renormalization is a
little more involved.\cite{LN0701,GLWtergV}  For an uniform
tensor-network on 2D square lattice (see Fig \ref{tnsq}), the
renormalization group procedure can be realized in two steps. The
first step is decomposing the rank-four tensor into two rank-three
tensors. We do it in two different ways on the sublattice purple and
green (see Fig.  \ref{CGsq}a).  On purple sublattice, we have
$T_{ruld}=\sum_{s=1}^{D^2} {S_1}_{uls}{S_3}_{drs}$ and on green
sublattice, we have $T_{ruld}=\sum_{s=1}^{D^2}
{S_2}_{lds}{S_4}_{rus}$.  Note that $r,l,u,d$ run over $D$ values
while $s$ run over $D^2$ values.  For such a range of $s$, the
decomposition can always be exact.

Next we try to reduce the range of $s$ through an
approximation.\cite{LN0701} Say, on purple sublattice, we view
$T_{ruld}$ as a matrix $M_{lu;rd}^{\rm{p}}= T_{ruld}$ and do singular
value decomposition
\begin{align}
\label{ULaV}
M^{\rm{p}}=U\La V^\dagger,\ \ \ \ \
\La=\bpm
\la_1 & & & \\
& \la_2 & & \\
& & &  \ddots\\
\epm
\end{align}
where $\la_1>\la_2>\la_3$ ...
We then keep only the largest $D_{cut}$ singular values $\la_s$
and define ${S_3}_{drs}=\sqrt{\la_s}V^\dag_{s,rd}$,
${S_1}_{uls} =\sqrt{\la_s}U_{lu,s}$.  Thus, we can approximately
express $T_{ruld}$ by two rank-three tensors $S_1$, $S_3$
\begin{eqnarray}
T_{ruld}\simeq\sum_{s=1}^{D_{\rm{cut}}}
{S_1}_{uls} {S_3}_{drs} .
\label{rule1}
\end{eqnarray}
Similarly, on green sublattice we may also define $T_{ruld}$ as a
matrix $M_{ld;ru}^{\rm{g}}$ and do singular value decompositions, keep
the largest $D_{\text{cut}}$ singular values and approximately express
$T_{ruld}$ by two rank-three tensors $S_2,S_4$.
\begin{eqnarray}
\c T_{\al\bt\mu\nu}\simeq\sum_{\ga=1}^{D_{\rm{cut}}}
{S_2}_{\nu\al\ga}{S_4}_{\bt\mu\ga}\label{rule2}
\end{eqnarray}

After such decompositions, the square lattice is deformed into the
form in Fig. \ref{CGsq}b.  The second step is simply contract the
square and get a new tensor $T'$ on the coarse grained lattice (see
Fig. \ref{tsrd}b).  The dimension for the reduced tensor $T'$ is only
$D_\text{cut}$ which can be chosen to be $D$ or some other values.
This defines the (approximated) renormalization flow of the tensor
network based on the SVD. Let us call such an approach SVD based
tensor renormalization group (SVDTRG) approach.

\section{The physical meaning of fixed-point tensors}

\label{cdlphy}

\begin{figure}
\begin{center}
\includegraphics[scale=0.6]
{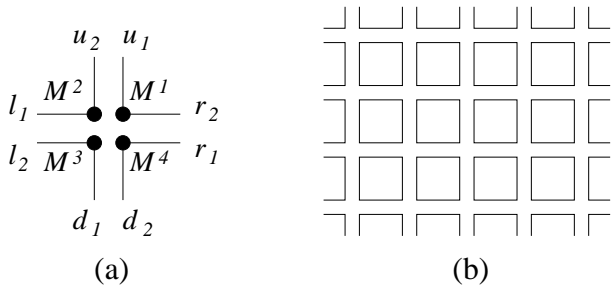}
\end{center}
\caption{
(a) A CDL tensor.
(b) A network of CDL tensor.
}
\label{cdl}
\end{figure}

Why the SVDTRG flow fails to produce a isolated fixed-point tensors?
Levin has pointed out that one large class of tensors -- corner
double-line (CDL) tensors -- are fixed points of the SVDTRG
flow.\cite{LN0701,L0795} A CDL tensor has a form (see Fig. \ref{cdl}a)
\begin{align}
\label{cdlts}
T_{ruld}(M^1,M^2,M^3,M^4) & = T_{r_1r_2,u_2u_1,l_2l_1,d_1d_2}
\nonumber\\
&= M^1_{r_2u_1} M^2_{u_2l_1} M^3_{l_2d_1} M^4_{d_2r_1}
\end{align}
where the pair $(r_1,r_2)$ represents the index $r$, \etc.  Such type
of tensor (assuming $\sqrt{D}=$ integer and $r_1$ \etc run from $1$ to
$\sqrt{D}$) has exact decomposition
\begin{align*}
T_{ruld} &= \sum_{s=1}^{D_\text{cut}} {S_3}_{drs}{S_1}_{lus}
 = \sum_{s=1}^{D_\text{cut}} {S_2}_{lds}{S_4}_{rus}
\end{align*}
with $D_\text{cut}=D$.  This makes the CDL tensor to be the
fixed-point tensor under the SVDTRG flow at least when
$M^1=M^2=M^3=M^4$.  If the initial tensor is changed slightly, the
tensor may flow to a slight different CDL
tensor, and hence the fixed point is not isolated.

\begin{figure}
\begin{center}
\includegraphics[scale=0.6]
{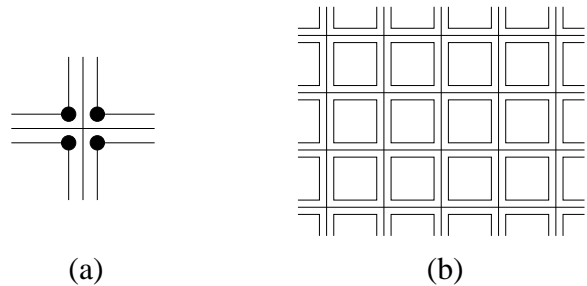}
\end{center}
\caption{
The fixed-point tensor and its tensor network
for topological phases.
}
\label{cdltop}
\end{figure}

A tensor network of CDL tensor is plotted in Fig.  \ref{cdl}b.  If the
system described by a tensor network does flow to a tensor network of
CDL tensor, what does this means? From  Fig.  \ref{cdl}b, we see that,
at the fixed point described by the CDL tensor, the degrees of freedom
form clusters that correspond to the disconnected squares.  Only the
degrees of freedom within the same square interact and there are no
interactions between different clusters. Such a fixed-point tensor
(\ie Lagrangian) describes a system with no interaction between
neighbors and has a finite energy gap.  So the system is in a trivial
phase without symmetry breaking and without topological order
(in the sense of long range entanglement).  The
$h\gg J$ phase of the transverse-field Ising model
\begin{equation}
\label{tIsing}
 H_\text{tIsing}= -\sum_{i} (J\si^x_i \si^x_{i+1} +h \si^z_i)
\end{equation}
is such a phase.  We can check that the tensor network that describes
the above transverse-field Ising model indeed flows to a network of
CDL tensor when $h\gg J$.

On the other hand, the low energy physics of such phases is really
trivial.  $J=0$ and $h=\infty$ limit, the partition function of the
transverse-field Ising model is described by a dimension-one tensor
\begin{align}
\label{trvts}
T^\text{TRI}_{ruld} = \del_{r,1} \del_{u,1} \del_{l,1} \del_{d,1}
\end{align}
Thus we expect that the fixed-point tensor for the  $h\gg J$ phase
should be the above trivial dimension-one tensor.  Unfortunately, the
renormalization flow generated by SVDTRG fails to produce such simple
fixed-point tensor.

The $h\ll J$ phase of the transverse-field Ising model is a symmetry
breaking phase with two degenerate ground states.  Under the SVDTRG
flow, the tensor network of the transverse-field Ising model will flow
to a tensor network described by a tensor of the form
\begin{align}
\label{dsumts}
 T= T_\text{cdl} \oplus T_\text{cdl}
\end{align}
where $T_\text{cdl}$ is a CDL tensor.  On the other hand, the symmetry
breaking phase should be described by a simpler tensor \eqn{dsumts}
where $T_{cdl}$ is the dimension-one tensor $T^\text{TRI}$ given by
\eqn{trvts}.

However, the SVDTRG flow does a remarkable job of squeezing all the
complicated entanglement into a single plaquette! Note that the
initial tensor network may contain entanglement between lattice sites
separated by a distance less than the correlation length. A CDL tensor
contains only entanglement between nearest-neighbor lattice sites,
while the ideal fixed-point tensor \eqn{trvts} has no entanglement at
all (not even with next neighbors).

\begin{figure}[tb]
\begin{center}
\includegraphics[scale=0.6]
{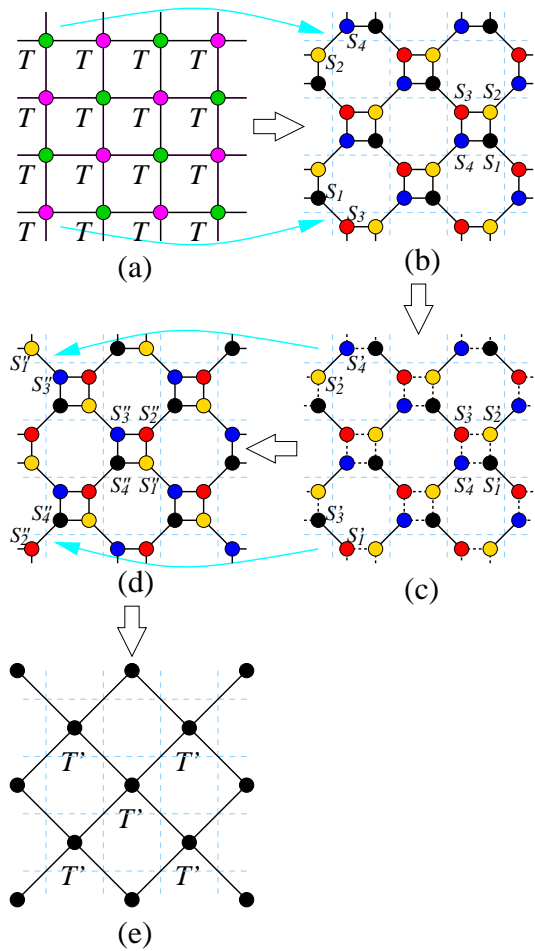}
\end{center}
\caption{
(Color online)
The TEFR procedure.
We first deform the tensor network
on square lattice (a) to the tensor network on lattice (b) using the
transformations in Fig. \ref{tsrd}(a).  We then replace the four
rank-three tensors $S_1,S_2,S_3,S_4$ in (b) by a new set of simpler
rank-three tensors $S'_1,S'_2,S'_3,S'_4$ in (c) while keeping the
tensor-trace of the tensor network approximately the same (see Fig.
\ref{S1234}).  The index on a solid link has a dimension $D$.  The
index on a dashed link has a dimension $D'$.  Thus the tensor $S_i$
has a dimension $D\times D\times D$  while the tensor $S'_i$ has a
dimension $D\times D\times D'$.  The SVD method is used to go from (a)
to (b) [see Fig. \ref{tsrd}(a)] and from (c) to (d) (see Fig.
\ref{StoS}).  The four $S''_i$ on the small squares in (d) are traced
over to obtain $T'$ in (e).
}
\label{TEFsqCG}
\end{figure}

What should be the fixed-point tensor for topological phases under
SVDTRG flow? The essence of topological phases is pattern of
long-range entanglement.\cite{Wqogen,KP0604,LWtopent} Since the CDL
tensor has no long range entanglement, its cannot be used to describe
topologically ordered phases.  We believe that topological phases are
described by fixed-point tensors of the following form (see Fig.
\ref{cdltop})
\begin{align}
\label{cdltopeq}
 T=T_{cdl}\otimes T_{top}
\end{align}
where $T_{cdl}$ is CDL tensor and $T_{top}$ is a tensor that cannot be
decomposed further as $T_{top}=T'_{cdl}\otimes T'_{top}$.  Here
$T_{top}$ will be the fixed-point tensor that characterizes the
topological order.

\begin{figure}[tb]
\begin{center}
\includegraphics[scale=0.8]
{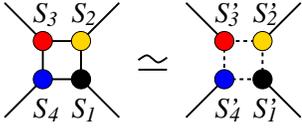}
\end{center}
\caption{
(Color online)
A tensor trace of a product of four rank-three tensors
$S_1,S_2,S_3,S_4$ give rise to one rank-four tensor.  Such a rank-four
tensor can also be produced (approximately) by a different set of
rank-three tensors $S'_1,S'_2,S'_3,S'_4$.  We may choose $S'_i$ trying
to minimize the dimension of the index on the dashed links.  Or we may
choose $S'_i$ trying to optimize some other quantities.
}
\label{S1234}
\end{figure}

\begin{figure}[tb]
\begin{center}
\includegraphics[scale=0.8]
{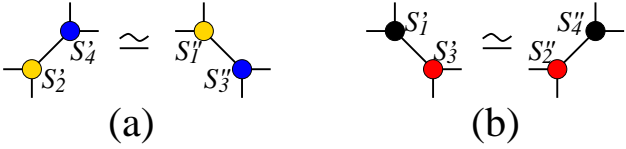}
\end{center}
\caption{
(Color online)
The SVD method is used to implement the above two transformations.
}
\label{StoS}
\end{figure}

\section{Filtering out the local entanglements}

We have seen that the SVDTRG flow can produce a network of CDL tensor
which squeezes the non-topological short range entanglement into
single plaquette.  To obtain ideal fixed-point tensor (such as the
one in \eqn{trvts}) we need to go one step further to remove the
corner entanglement represented by the CDL tensor, \ie we need to
reduce corner matrix $M^i$ in the CDL tensor (see \eqn{trvts}) to a
simpler form
\begin{align}
M^i_{ab}\to \del_{a,1}\del_{b,1}.
\end{align}
In the following, we will describe one way to do so.

\begin{figure}[tb]
\begin{center}
\includegraphics[scale=0.8]
{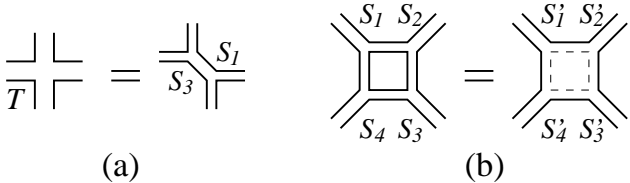}
\end{center}
\caption{
TEFR transformation of a CDL tensor.
Here the index on a dash-line has a dimension one.
}
\label{TScdl}
\end{figure}

\begin{figure}[tb]
\begin{center}
\includegraphics[scale=0.55]
{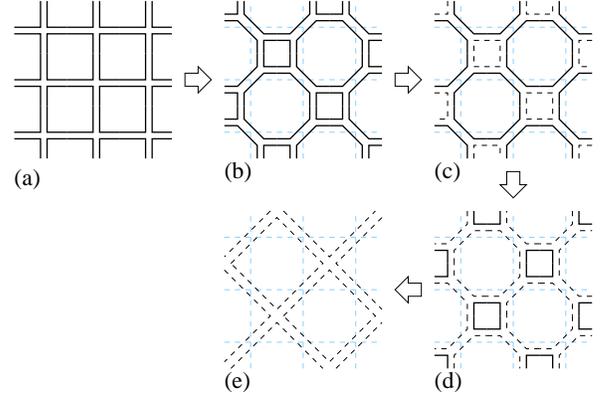}
\end{center}
\caption{
A tensor network of CDL tensor can be reduced to a tensor network of
dimension-one tensor.
(a) $\to$ (b) uses the transformation in Fig. \ref{TScdl}(a).
(b) $\to$ (c) uses the transformation in Fig. \ref{TScdl}(b).
(c) $\to$ (d) uses the transformation in Fig. \ref{StoS}.
(d) $\to$ (e) uses the transformation in Fig. \ref{CGsq}(b)$\to$(c).
Here the index on a dash-line has a dimension one.
}
\label{cdltri}
\end{figure}

In this approach, we modify the SVDTRG transformation described by
Fig. \ref{CGsq} to a new transformation described by Fig.
\ref{TEFsqCG}.  We will call the new approach
tensor-entanglement-filtering renormalization (TEFR) approach.  In the TEFR
transformation, we first deform the square lattice Fig.
\ref{TEFsqCG}(a) into the lattice Fig. \ref{TEFsqCG}(b) by expressing
the rank-four tensor $T$ in terms of two rank-three tensors $S_1,S_3$
or $S_2,S_4$ [see Fig.  \ref{tsrd}(a)].  Then we replace the four
rank-three tensors $S_1,S_2,S_3,S_4$ by a new set of simpler
rank-three tensors $S'_1,S'_2,S'_3,S'_4$ so that the tensor trace of
$S'_1,S'_2,S'_3,S'_4$ (approximately) reproduces that of
$S_1,S_2,S_3,S_4$ as shown in Fig. \ref{S1234}.  In this step, we try
to reduce the dimensions of the rank-three tensors, from $ D\times
D\times D$ of $S_i$ to $ D\times D\times D'$ of $S'_i$ with $D'<D$.
If $T$ is a CDL tensor, $S_1,S_2,S_3,S_4$ will also be CDL tensors
[see Fig. \ref{TScdl}(a)].  In this case, the above procedure will
result in a new set of tensors $S'_1,S'_2,S'_3,S'_4$ with a lower
dimension as shown in Fig.  \ref{TScdl}(b).
(For details, see Appendix \ref{teftnr}.)  We then use the SVD (see
Fig. \ref{StoS}) to deform the lattice in Fig. \ref{TEFsqCG}(c) to
that in Fig. \ref{TEFsqCG}(d).  Last we trace out the indices on the
small squares in Fig.  \ref{TEFsqCG}(d) to produce new tensor network
on a coarse grained square lattice [see \ref{TEFsqCG}(e)] with a new
tensor $T'$.  The tensor $T'$ will have a lower local entanglement
than $T$.  In fact, when $T$ is a CDL tensor, the combined procedures
described in Fig.  \ref{CGsq} and Fig.  \ref{TEFsqCG} (see also Fig.
\ref{TScdl} and \ref{cdltri}) can reduce $T$ to a dimension-one tensor.
Appendix \ref{teftnr} contains a more detailed description
of TEFR method.

\section{Examples}

In the section, we will apply the TEFR method to study several simple
well known statistical and quantum systems to test the TEFR approach.
We will calculate the free energy (for statistical systems) or the
ground-state energy (for quantum systems).  The singularities in
free-energy/ground-state-energy in the large system limit will
indicate the presence of phase transitions.  We can also calculate
the fixed-point tensors under the TEFR flow for each phase. The
fixed-point tensors will allow us to identify those phases as symmetry
breaking and/or topological phases.  If we find a continuous
phase transition in the above calculation, we can also use the
fixed-point tensor at the critical point to calculate the spectrum of
the scaling dimensions $h_i$ and the central charge $c$ for the
critical point (see Appendix \ref{partcri}).

\begin{figure}[tb]
\begin{center}
\includegraphics[scale=1.]
{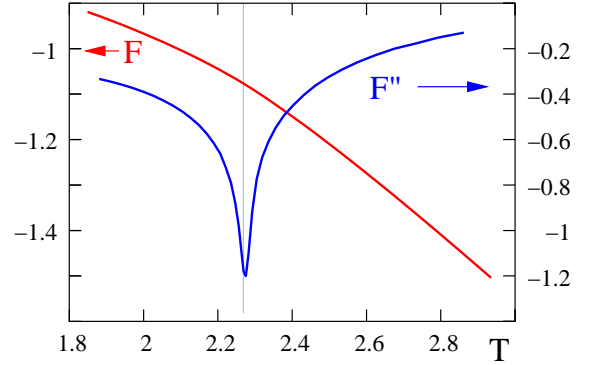}
\end{center}
\caption{
(Color online) The temperature dependence of free energy $F(T)$ and
$F''=\frac{\prt^2 F}{\prt T^2}$ for the 2D statistical Ising model.
A phase transition can be seen around the temperature $T\approx 2.3$.
The vertical line marks the exact $T_c=2.26919$.
}
\label{freeIsing}
\end{figure}

\subsection{2D statistical Ising model}

Our first example is the 2D statistical Ising model on square lattice
described by
\begin{equation}
\label{ising}
 H=-\sum_{\<\v i,\v j\>}
\si_{\v i} \si_{\v j},\ \ \ \ \ \si_{\v i}=\pm 1
\end{equation}
where $\sum_{\<\v i,\v j\>} $ sums over nearest neighbors.  The
partition function is given by $Z=\Tr(e^{-\beta H})$.  Such a
partition function can be expressed as a tensor-trace over a
tensor-network on a (different) square lattice where the tensor $T$ is
given by \begin{align} \label{tnsIsing}
T^\text{Ising}_{1,2,1,2}&=e^{-4\bt}, &
T^\text{Ising}_{2,1,2,1}&=e^{-4\bt}, \nonumber\\
T^\text{Ising}_{1,1,1,1}&=e^{4\bt}, &
T^\text{Ising}_{2,2,2,2}&=e^{4\bt}, \nonumber\\ \text{others} &= 1.
\end{align} Note that the spins are located on the links of the
tensor-network (\ie $\v i$ in \eq{ising} labels the links of the
tensor-network).

The statistical Ising model has a $Z_2$ spin flip symmetry.
Such a symmetry implies that the tensor is invariant under the
following $Z_2$ transformation
\begin{align}
\label{Z2Ising}
 T^\text{Ising}_{ruld}= W_{rr'} W_{uu'} W_{ll'} W_{dd'}
 T^\text{Ising}_{r'u'l'd'},
\end{align}
where \begin{align*}
 W=\bpm
0 & 1 \\
1 & 0 \\
\epm  .
\end{align*}

We performed 14 steps of TEFR iteration described in Fig. \ref{TEFsqCG}
with $D_\text{cut}=32$.  The calculated free energy $F=-\bt^{-1} \log (Z)
=-T \log (Z)$ is presented in Fig. \ref{freeIsing}.  For temperature
$T>2.38$, the fixed-point tensor is a dimension-one tensor
\begin{eqnarray}
\label{T0}
T^\text{TRI}_{1111}=1,  \ \ \ \ \ \ \
\text{others} = 0.
\end{eqnarray}
For temperature $T<2.19$, the fixed-point tensor is a
dimension-two tensor
\begin{eqnarray}
\label{Z2ts}
T^{Z_2}_{1111}=1,  \ \ \ \ \ \ \
T^{Z_2}_{2222}=1,  \ \ \ \ \ \ \
\text{others} = 0.
\end{eqnarray}
For temperature $2.22<T< 2.32$, the fixed-point tensor is complicated
with many non-zero elements.  (We like to point out that the above
fixed-point tensors are actually invariant fixed-point tensors.  For
details, see Appendix \ref{partcri}.)

\begin{figure}[tb]
\begin{center}
\includegraphics[scale=0.5]
{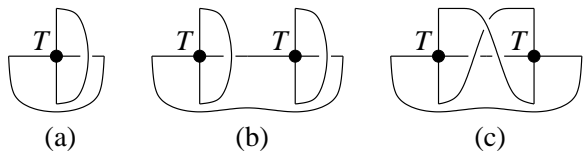}
\end{center}
\caption{
The graphic representations of (a)
$\sum_{ru} T_{ruru}$, (b)
$\sum_{ruld} T_{rulu}T_{ldrd}$, and (c)
$\sum_{ruld} T_{ruld}T_{ldru}$.
}
\label{X1X2}
\end{figure}

\begin{figure}[tb]
\begin{center}
\includegraphics[scale=1.]
{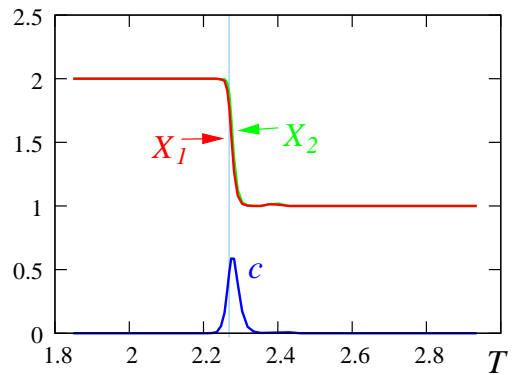}
\end{center}
\caption{
(Color online) The temperature dependence of
$X_1$, $X_2$, and the central charge $c$
for the 2D statistical Ising model.
The vertical line marks the exact $T_c=2.26919$.
}
\label{X12cIsing}
\end{figure}

To visualize the structure of fixed-point tensors more
quantitatively in different regions,
we introduce the following two quantities (see Fig. \ref{X1X2})
\begin{align}
\label{X12}
 X_1&=\frac{(\sum_{ru} T_{ruru})^2}{\sum_{ruld} T_{rulu}T_{ldrd}} ,
&
 X_2&=\frac{(\sum_{ru} T_{ruru})^2}{\sum_{ruld} T_{ruld}T_{ldru}}
\end{align}
for the fixed-point tensors.  Note that $X_1$ and $X_2$ is independent
of the scale of the tensor $T\to \Ga T$.  We plotted $X_1$, $X_2$ and
the central charge obtained from \eq{chla} for the
fixed-point tensors $T_\text{inv}$ at different temperatures in Fig.
\ref{X12cIsing}. (Note that $\tau_0=\tau_1=\imth$ here.) We see that the
different values of $X_1$ and $X_2$ mark the different phases, and the
steps in  $X_1$ and $X_2$, as well as the peak in $c$ mark the point
of continuous transition.

From the structure of the fixed-point tensor, we see that the
low-temperature phase $T<2.19$ is a $Z_2$ symmetry breaking phase
(where the fixed-point tensor has a form $T^{Z_2}=T^\text{TRI}\oplus
T^\text{TRI}$) and the high-temperature phase $T>2.38$ is a trivial
disordered phase.
Since $\prt F/\prt T$ appear to be continuous and $\prt^2 F/\prt T^2$
appears to have an algebraic divergence at the transition point, all
those suggest that the phase transition is a continuous transition,
which is a totally expected result.  From the Onsager's solution\cite{O4417}
$\text{sinh}(2/T_c)=1$, one obtains the exact critical temperature to
be $T_c=\frac{2}{\log(\sqrt 2+1)} =2.26919$, which is consistent with
our numerical result.

There is another interpretation of the fixed-point tensors.  The
fixed-point in the low temperature phase $T^{Z_2}=T^\text{TRI}\oplus
T^\text{TRI}$ happen to be the initial tensor \eq{tnsIsing} in the zero
temperature limit $\bt=\infty$.  Thus the flow of the tensor in the
low temperature phase can be viewed as a flow of the temperature
$\bt\to \infty$.  Using the same reasoning, we expect that the
flow of the tensor in the high temperature phase can also be viewed as
a flow of temperature $\bt\to 0$ in the opposite direction.
This would suggest that the fixed-point tensor should have a form
\begin{equation}
\label{TTRI}
 \t T^\text{TRI}_{ruld}=\frac{1}{4}, \ \ \ \ r,u,l,d=1,2
\end{equation}
which is very different from the trivial dimension-one tensor
$T^\text{TRI}$ in \eq{T0}.  We will explain below that the two tensors
$\t T^\text{TRI}$ and $T^\text{TRI}$ are equivalent.

To see the equivalence,
we would like to point out that two tensors $T$ and $T'$ related by
\begin{align}
\label{fred}
 T'_{r'u'l'd'}=
(A^{-1})_{l'l}
(B^{-1})_{u'u}
T_{ruld}
A_{rr'}
B_{dd'}
\end{align}
give rise to same tensor trace for any square-lattice tensor  network.
In particular, if the singular values in the SVD decomposition in Fig.
\ref{tsrd}(a) has degeneracies, the coarse grained tensor will have
the ambiguity described by the above transformation with $A$ and $B$
being orthogonal matrices.  Physically, the transformation \eq{fred}
corresponds to a field redefinition.
One can check
explicitly that
\begin{equation*}
\t T^\text{TRI}_{ruld}=
R^{-1}_{rr'} R^{-1}_{uu'} T^\text{TRI}_{r'u'l'd'} R_{l'l} R_{d'd}
\end{equation*}
where
\begin{equation}
\label{Rmat}
 R=\frac{1}{\sqrt 2} \bpm 1  & 1\\
                          1  & -1 \epm
\end{equation}
Thus, $\t T^\text{TRI}$ and $T^\text{TRI}$ are equivalent under a
field redefinition.

We have seen that the coarse grained tensors after the TEFR
transformations may contain a field redefinition ambiguity \eq{fred}.
As a result, the $Z_2$ transformation on the coarse grained lattice
may be different from that on the original lattice.  The $Z_2$
transformation on the coarse grained lattice is related to the
original $Z_2$ transformation by a field redefinition transformation.

However, one can choose a basis on the coarse grained lattice by doing
a proper field-redefinition transformation such that the $Z_2$ spin
flip transformation has the same form on the coarse grained lattice as
that on the original lattice.  In such a basis, the disordered and the
$Z_2$ symmetry breaking phases are described by
$(G_\text{sym},T_\text{inv}) = (Z_2,\t T^\text{TRI})$ and
$(G_\text{sym},T_\text{inv}) = (Z_2, T^{Z_2})$, respectively, where
the $Z_2$ transformation is given by \eq{Z2Ising}.  This example shows
how to use the pair $(G_\text{sym},T_\text{inv})$ to describe the two
phases of the statistical Ising model.

Next we like to use the fixed-point tensor $T_\text{inv}$ at the
critical point to calculate the critical properties of the above
continuous transition.  We can use \eq{Tinv} to obtain
the invariant fixed-point tensor
$T^{(i)}_\text{inv}$ even off the critical point.  From
$T^{(i)}_\text{inv}$ we calculate the matrices $M^{ud}$, $M^{lr}$,
$M^{ldru}$, and $M^{lurd}$ from \eq{TM}, and obtain the eigenvalues
$\la_n$ of those matrices.  Using those eigenvalues, we can calculate
the central charge and the scaling dimensions from \eq{chla}, where
$\tau=\tau_1=\imth$ for the Ising model discussed here.  By
choosing $D_\text{cut}=64$ and doing 9 iterations of Fig.
\ref{TEFsqCG} (that correspond to a system with 1024 spins)
at the critical temperature $T_c=\frac{2}{\log(\sqrt 2+1)}$, we find
that
\begin{align*}
& c && h_1 && h_2 && h_3 && h_4
\nonumber\\
& 0.49942 && 0.12504 && 0.99996 && 1.12256 && 1.12403
\nonumber\\
& 1/2 && 1/8 && 1 && 9/8 && 9/8
\end{align*}
where we also listed the exact values of the central charge and the
scaling dimensions.  We would like to mention that the TEFR approach
does not suppose to work well at the critical point where the
truncation error of keeping only a finite $D_\text{cut}$ singular
values is large.  It turns out that for $D_\text{cut}=64$, the
estimated relative truncation error is less than $10^{-5}$ through $9$
iterations. This allows us to obtain quite accurate central charge and
scaling dimensions.  The truncation error will grow for more
iterations. As a result, the agreement with the exact result will
worsen.

From this example, we see that the TEFR approach is an effective and
efficient way to study phases and phase transitions.
The TEFR approach allow us to identify spontaneous symmetry breaking
from the structure of fixed-point tensors.  The critical properties
for the continuous phase transition can also be obtained from the
fixed-point tensor at the critical point.

\subsection{2D statistical loop-gas model}

Next we consider a 2D statistical loop-gas model on square lattice.
The loop-gas model can be viewed as a Ising model
with spins on the links of the square lattice.
However, the allowed spin configurations
must satisfies the following hard constraint:
the number of up-spins next each vertex must be even.
The energy of spin configuration is
given by
\begin{equation}
\label{loopgas}
 H=-\sum_{\<\v i\>} \si_{\v i}
\end{equation}
The partition function $Z=\Tr(e^{-\beta H})$ can be expressed as a
tensor-trace over a tensor-network on the square lattice where the
tensor $T$ is given by
\begin{align}
\label{tnslg}
T^\text{LG}_{ruld} &=
\e^{\bt (2 d-3)/2}
\e^{\bt (2 u-3)/2}
\e^{\bt (2 r-3)/2}
\e^{\bt (2 l-3)/2}\times
\nonumber\\
&\ \ \ \ \ \  \del_{d+u+r+l=\text{even}}
\end{align}
where the indices take the values
$r,u,l,d=1,2$ and
\begin{equation*}
  \del_{d+u+r+l=\text{even}}=1 \ \ \text{ if }d+u+r+l=\text{even},\ \
\ \ \text{others}=0 .
\end{equation*}
To use the TEFR approach to calculate the partition
function, it is very important to implement the TEFR in
such a way that the closed-loop condition
\begin{align}
\label{loopT}
 T_{ruld}&=0, \text{ when } r+u+l+d=\text{odd}
\nonumber\\
 S_{rul}&=0, \text{ when } r+u+l=\text{odd}
\end{align}
are satisfied at every step of iteration.

\begin{figure}[tb]
\begin{center}
\includegraphics[scale=1.]
{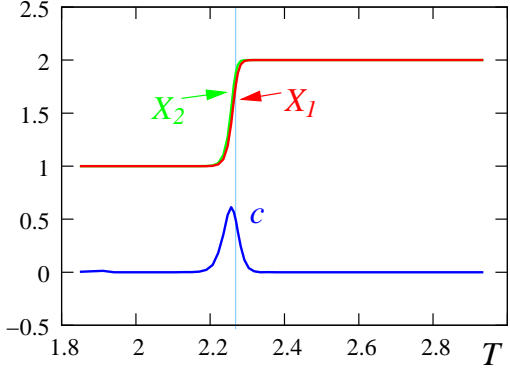}
\end{center}
\caption{
(Color online) The temperature dependence of
$X_1$, $X_2$, and the central charge $c$
for the 2D statistical loop gas model.
The vertical line marks the exact $T_c=2.26919$.
}
\label{X12cloop}
\end{figure}

We performed 14 steps of TEFR iteration described in Fig. \ref{TEFsqCG}
with $D_\text{cut}=32$.  Fig.  \ref{X12cloop} shows the resulting
$X_1$, $X_2$, and the central charge $c=\frac{6}{\pi}\log(\la_0)$
obtained from \eq{chla} for the invariant fixed-point tensors
$(T_\text{inv})_{ruld}$ at different temperatures.  We find that the
low temperature phase is described by the trivial dimension-one
fixed-point tensor $T^\text{TRI}$ \eq{trvts} as indicated by
$X_1=X_2=1$.  The high temperature phase has a non-trivial fixed-point
tensor since  $X_1=X_2=2$.  We find that the invariant fixed-point
tensor to have a form
\begin{align}
\label{Tll}
 T^\text{LL}_{ruld}&=\frac{1}{2}, \text{ if } l+r+u+d=\text{even},\ \
r,u,l,d=1,2
\nonumber\\
\text{others}&=0.
\end{align}
The two fixed-point tensors $T^\text{TRI}$ and $T^\text{LL}$
correspond to the zero-temperature and infinite-temperature limit of
the initial tensor $T^\text{LG}$ in \eq{tnslg}.  Both fixed-point
tensors satisfy the closed-loop condition \eq{loopT}.

It is well known that the high temperature phase of the statistical loop gas
model is dual to the low temperature phase of the statistical Ising
model, and the low temperature phase of the statistical loop gas model
is dual to the high temperature phase of the statistical Ising model.
The fixed-point tensors of the two models are related directly
\begin{align*}
T^\text{LL}_{ruld} &=
R^{-1}_{rr'} R^{-1}_{uu'} T^{Z_2}_{r'u'l'd'} R_{l'l} R_{d'd}
\nonumber\\
T^\text{TRI}_{ruld} &=
R^{-1}_{rr'} R^{-1}_{uu'} \t T^\text{TRI}_{r'u'l'd'} R_{l'l} R_{d'd}
\end{align*}
where $R$ is given in \eq{Rmat}.  The critical properties at the continuous
transition are also identical.

One may wonder, if $T^\text{LL}$ and $T^{Z_2}$ are related by a field
redefinition, then why the TEFR flow produces $T^\text{LL}$ as the
fixed-point tensor of the high temperature phase of the loop gas,
rather than $T^{Z_2}$?  Indeed, if we implement the TEFR flow in the
space of generic tensors, then both $T^\text{LL}$ and $T^{Z_2}$,
together with many other tensors related by some field redefinitions
will appear as the fixed-point tensors of the high temperature phase
of the loop gas.  However, if we implement the TEFR flow in the space
of tensors that the describe closed loops (\ie satisfying \eq{loopT}),
then only $T^\text{LL}$ will appear as the fixed-point tensor.

\subsection{Generalized 2D statistical loop-gas model}

Third, let us discuss a more general 2D statistical loop-gas model described by
a tensor network with tensor
\begin{align}
\label{tnsGLG}
T^\text{GLG}_{ruld} &=
\e^{ -(r+l+u+d-6)\bt + V \del_{rlud}} \del_{d+u+r+l=\text{even}} ,
\end{align}
where $r,u,l,d=1,2$ and $\del_{rlud}$ is given by
\begin{equation*}
 \del_{1111}=\del_{2222}=1,\ \ \ \ \text{others}=0.
\end{equation*}
We note that when $V=0$, the generalized loop-gas model become the one
discussed above: $T^\text{LG}=T^\text{GLG}|_{V=0}$.

Let us consider the phases of the generalized loop-gas model along the
$\bt=0$ line.  In this case, the loop-gas model also has a $Z_2$
symmetry: the tensor $T^\text{GLG}$ is invariant under switching the 1
and 2 index:
\begin{align}
\label{Z2x}
 T^\text{GLG}_{ruld} &= W_{rr'} W_{uu'} W_{ll'} W_{dd'}
 T^\text{GLG}_{r'u'l'd'},
\nonumber\\
 W_x &=\bpm
0 & 1 \\
1 & 0 \\
\epm  .
\end{align}
We will call such a symmetry $Z^x_2$ symmetry.
The closed-loop condition \eq{loopT} can also be represented as a
$Z_2$ symmetry of the tensor:
\begin{align}
\label{Z2z}
 T^\text{GLG}_{ruld} &= W_{rr'} W_{uu'} W_{ll'} W_{dd'}
 T^\text{GLG}_{r'u'l'd'},
\nonumber\\
 W_z &=\bpm
1 & 0\\
0 &-1 \\
\epm  .
\end{align}
The second $Z_2$ symmetry will be called $Z^z_2$ symmetry.  So along
the $\bt=0$ line, the model has a $Z^x_2\times Z^z_2$ symmetry.

\begin{figure}[tb]
\begin{center}
\includegraphics[scale=1.]
{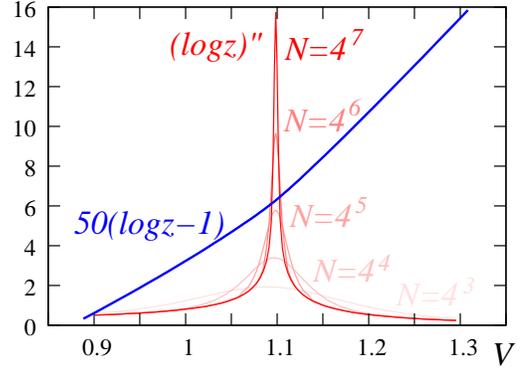}
\end{center}
\caption{
(Color online)
The $V$ dependence of $logz=\log (Z)/N$ and $(logz)"=d^2 logz/d V^2$
for the generalized 2D loop gas model.
The sharpest peak for $(logz)"$ and the $logz$ currve
are for a tensor network with $N=4^7$
tensors.  Other peaks are for tensor networks with $N=4^6$, $N=4^5$,
$N=4^4$, and $N=4^3$ tensors.
}
\label{logzGLG}
\end{figure}

\begin{figure}[tb]
\begin{center}
\includegraphics[scale=1.]
{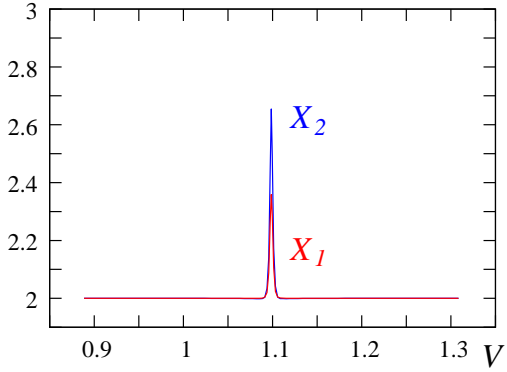}
\end{center}
\caption{
(Color online) The $V$ dependence of $X_1$, $X_2$
for the generalized 2D loop gas model.
}
\label{X12GLG}
\end{figure}

\begin{figure}[tb]
\begin{center}
\includegraphics[scale=1.]
{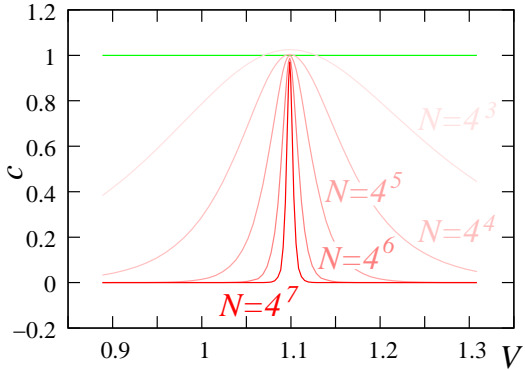}
\end{center}
\caption{
(Color online) The $V$ dependence
of the central charge $c$
for the generalized 2D loop gas model.
The sharpest peak is for a tensor network with $N=4^7$
tensors.  Other peaks are for tensor networks with $N=4^6$, $N=4^5$,
$N=4^4$, and $N=4^3$ tensors.
}
\label{cGLG}
\end{figure}

We performed the TEFR calculation for the generalized loop-gas model
along the $\bt=0$ line with $D_\text{cut}=32$.  Fig. \ref{logzGLG}
shows the $V$ dependence of the resulting log partition function per
tensor, $logz=log(Z)/N$, and $d^2logz/dV^2$.  Fig. \ref{X12GLG} and
Fig. \ref{cGLG} show the $V$ dependence of $X_1$, $X_2$, and central
charge $c$ for the resulting fixed-point tensor.  We see that there is
a phase transition at $V_c=1.0985$.  The fixed-point tensor for the
$V<V_c$ phase is found to be $T^\text{LL}$ and the fixed-point tensor
for the $V>V_c$ phase is found to be $T^{Z_2}$.  The central charge is
zero for the two phases on the two sides of the transition.  Thus the
both phases have short range correlations.  Using the
$(G_\text{sym},T_\text{inv})$ notation, the two phases are
characterized by $(Z^x_2\times Z^z_2,T^\text{LL})$ and $(Z^x_2\times
Z^z_2,T^{Z_2})$.

Since $\prt \log(Z)/\prt V$ has no discontinuous jump
at $V_c$, the transition is a continuous
phase transition.  Also the central charge is non zero at the
transition point which suggests that the correlation length diverges
at $V_c$.  This also implies that the transition is a continuous phase
transition. From Fig. \ref{cGLG}, we see that the transition is a
central charge $c=1$ critical point.

We note that the two fixed-point tensors $T^\text{LL}$ and $T^{Z_2}$
give rise to the same $X_1$ and $X_2$.  In fact, $T^\text{LL}$ and
$T^{Z_2}$ are related by a field-redefinition transformation
\eq{fred}:
\begin{equation*}
T^\text{LL}_{ruld} =
R^{-1}_{rr'} R^{-1}_{uu'} T^{Z_2}_{r'u'l'd'} R_{l'l} R_{d'd}
\end{equation*}
where $R$ is given in \eq{Rmat}.  Since $X_1$ and $X_2$ are invariant
under the above transformation, so they are the same in the two
phases.

The generalized loop-gas model provides us an interesting example that
the fixed-point tensors $T^\text{LL}$ and $T^{Z_2}$ that describe the
two phases are related by a field redefinition transformation. In this
case, one may wonder should we view the $V<V_c$ and $V>V_c$ phases as
the same phase? A simple direct answer to the above question is no.
It is incorrect to just use a fixed-point tensors to characterize a
phase. We should use the pair $(G_\text{sym},T_\text{inv})$ to
characterize a phase.  If we use the $(G_\text{sym},T_\text{inv})$
notation, the two phases are more correctly described by $(Z^x_2\times
Z^z_2,T^\text{LL})$ and $(Z^x_2\times Z^z_2,T^{Z_2})$.  The above
field-redefinition transformation transforms $T^\text{LL}$ to
$T^{Z_2}$.  It also transform the symmetry transformations
$Z^x_2\times Z^z_2$ to a different form $\t Z^x_2\times \t Z^z_2$.
Thus the pair $(Z^x_2\times Z^z_2,T^\text{LL})$ is transformed  to
$(\t Z^x_2\times \t Z^z_2,T^{Z_2})$ under the above field-redefinition
transformation.  Therefore, we cannot transform the pair $(Z^x_2\times
Z^z_2,T^\text{LL})$ to $(Z^x_2\times Z^z_2,T^{Z^2})$ using any field
redefinition transformation.  This implies that the two phases
$(Z^x_2\times Z^z_2,T^\text{LL})$ and $(Z^x_2\times Z^z_2,T^{Z_2})$
are different if we do not break the $Z^x_2\times Z^z_2$ symmetry.  If
we do break the $Z^x_2\times Z^z_2$ symmetry, then the two phases are
characterized by $(1,T^\text{LL})$ and $(1,T^{Z_2})$, where $1$
represents the trivial group with only identity implying that the
system has no symmetry.  Without any symmetry, the two labels
$(1,T^\text{LL})$ and $(1,T^{Z_2})$ are related under a field
redefinition transformation.  This implies that the two phases
$(1,T^\text{LL})$ and $(1,T^{Z_2})$ are the same phase if the system
has no symmetry.  In Appendix \ref{phase}, we will give a more
detailed and general discuss about the definition of phases and its
relation to the symmetry of the system.

\subsection{Quantum spin-1/2 chain and phase transition beyond
symmetry breaking paradigm}
\label{sp12}

The generalized loop-gas model is closely related to the following 1D
quantum spin-1/2 model:
\begin{equation}
\label{Hspin12}
H= \sum_i
[-\si^x_i\si^x_{i+1} -J \si^z_i\si^z_{i+1} - h \si^z_i ]
\end{equation}
Note that $\sum (1-\si^z_i)/2 =\sum_i n_i$ mod 2 is a
conserved quantity and the above model can be viewed as a 1D hard-core
boson model with next neighbor interaction.  Such a 1D quantum model
can be simulated by a 2D statistical loop-gas model described by a
tensor network with tensor $T^\text{GLG}$ \eq{tnsGLG}.  The loops in
the loop gas correspond to the space-time trajectories of the
hard-core bosons.  The $\bt=0$ limit of the loop-gas model correspond
to $h=0$ limit of the 1D quantum model, and $V$ corresponds to $J$.

In the following, we will concentrate on the $h=0$ case.  When $h=0$,
the Hamiltonian \eq{Hspin12} has a $Z^{x}_2\times Z^{z}_2$ symmetry,
where $Z^{x}_2$ is the $Z_2$ group generated by $\si^x$ and $Z^{z}_2$
is the $Z_2$ group generated by $\si^z$.  Such a $Z^{x}_2\times
Z^{z}_2$ symmetry corresponds to the $Z^{x}_2\times Z^{z}_2$ symmetry
discussed in last section.

When $J <J_c= 1$, the system is in a symmetry breaking phase. We will
use a pair of groups $(G_\text{sym},G_\text{grnd})$ to label such a
symmetry breaking phase, where $G_\text{sym}$ is the symmetry group of
the Hamiltonian and $G_\text{grnd}$ is the symmetry group of a ground
state.  So the $J <J_c$ phase of our system is labeled by
$(G_\text{sym},G_\text{grnd})=(Z^{x}_2\times Z^{z}_2,Z^{x}_2)$, where
the ground state breaks the $Z^{z}_2$ symmetry but has the $Z^{x}_2$
symmetry.  When $J > J_c$, our system is in a different symmetry
breaking phase, which is labeled by
$(G_\text{sym},G_\text{grnd})=(Z^{x}_2\times Z^{z}_2,Z^{z}_2)$.

Under the correspondence between the loop gas and our quantum spin-1/2
model, the $V<V_c$ and $V>V_c$ phases of the loop-gas model discussed
above will correspond to the above two $Z_2$ symmetry breaking phases
for $J<J_c=1$ and $J>J_c$ respectively.  The results from the loop gas
imply that the two $Z_2$ symmetry breaking phases are connected by a
\emph{continuous} phase transition.  Such a continuous phase transition
can be viewed as two $Z_2$ symmetry breaking transition happening at
the same point.

Strictly speaking, the above continuous transition does not fit within
the standard symmetry breaking paradigm for continuous phase
transitions.  According to Landau symmetry breaking theory, a
continuous phase transition can only happen between two phases labeled
by $(G_\text{sym},G_\text{grnd})$ and $(G_\text{sym},G'_\text{grmd})$
where $G_\text{grnd}$ is a subgroup of $G_\text{grnd}'$ or
$G_\text{grnd}'$ is a subgroup of $G_\text{grnd}$.  We see that the
continuous transition between the $(Z^{x}_2\times Z^{z}_2,Z^{x}_2)$
phase and $(Z^{x}_2\times Z^{z}_2,Z^{z}_2)$ phase does not fit within
the standard Landau symmetry breaking theory.  Such a continuous
transition is another example of continuous transitions that is beyond
the Landau symmetry breaking
paradigm.\cite{WWtran,CFW9349,SMF9945,RG0067,Wctpt,Wqoslpub,SBS0407}
Even though the above continuous phase transition does not fit within
the standard symmetry breaking theory, such a transition can still be
studied through the tensor network renormalization approach.  In
particular, the tensor network renormalization approach reveals that
the transition is a central charge $c=1$ critical point.

\begin{figure}[tb]
\begin{center}
\includegraphics[scale=.6]
{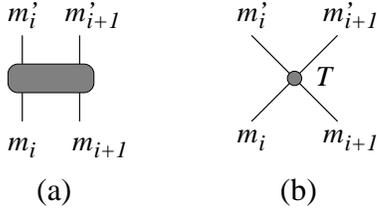}
\end{center}
\caption{
(a) The graphic representation
of $(\e^{-\del \tau H_i})_{m_im_{i+1},m'_im'_{i+1}}$.
(b) $\e^{-\del \tau H_i}$ can be viewed as a rank-four tensor.
}
\label{H2}
\end{figure}

\subsection{Quantum spin-1 chain}

Last, we consider a quantum spin-1 chain at zero temperature.  The
Hamiltonian is given by
\begin{equation}
\label{spin1}
 H=\sum_{i} \left( \v S_{i}\cdot \v S_{i+1} + U (S^z_{i})^2
+ B S^x_i
\right)
\end{equation}
The model has a translation symmetry,
a $Z_2$ time reversal symmetry $S^y\to -S^y$ and
$Z_2$ parity symmetry.  The imaginary-time
path-integral of the model can be written as
\begin{align*}
 \Tr \e^{-\tau H}=\lim_{M\to \infty}
 \Tr (\
e^{-\del\tau H_e}
e^{-\del\tau H_o}
)^M
\end{align*}
where $\del \tau=\tau/M$ and
\begin{align*}
 H_e &=\sum_{i=\text{even}} \Big(
\v S_{i}\cdot \v S_{i+1} + U (S^z_{i})^2
+ \frac{B}{2} (S^x_i+S^x_{i+1})
\Big)
\nonumber\\
 H_o &=\sum_{i=\text{odd}} \Big(
\v S_{i}\cdot \v S_{i+1} + U (S^z_{i})^2
+ \frac{B}{2} (S^x_i+S^x_{i+1})
\Big)
\end{align*}

\begin{figure}[tb]
\begin{center}
\includegraphics[scale=.48]
{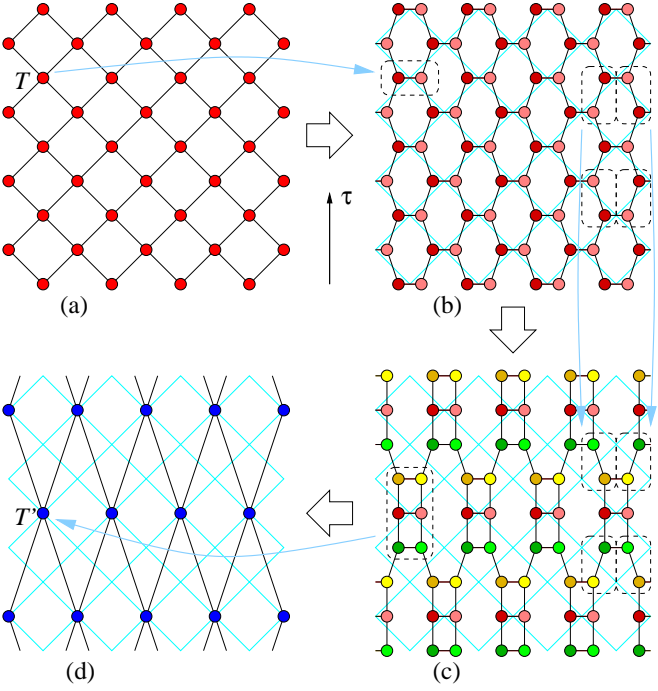}
\end{center}
\caption{
(Color online)
A $T_a$-$T_b$ tensor network of size $L\times M$ is reduced
to a $T'_a$-$T'_b$ tensor network of size $L\times M/2$.
From (a) to (b), the local deformation Fig. \ref{tsrd}(a) is used.
From (b) to (c), the local deformation Fig. \ref{StoS} is used.
From (c) to (d), a local deformation similar to that
in Fig. \ref{tsrd}(b) is used.
}
\label{stepY}
\end{figure}

Note that $H_e$ and $H_o$ are sum of non overlapping terms and thus
\begin{align*}
\e^{-\del\tau H_e} &=\prod_{i=\text{even}} \e^{-\del\tau H_i}
\nonumber\\
\e^{-\del\tau H_o} &=\prod_{i=\text{odd}} \e^{-\del\tau H_i} .
\end{align*}
$H_i$ in the above expression has a form
\begin{align*}
 H_i=\v S_i\cdot \v S_{i+1} + U (S^z_i)^2
+ \frac{B}{2} (S^x_i+S^x_{i+1})
\end{align*}
The matrix elements of $\e^{-\del\tau H_i}$
can be represented by a rank-four tensor (see Fig. \ref{H2})
\begin{align}
\label{THi}
 T_{ m_{i+1}, m_i, m'_i, m'_{i+1} }
=(\e^{-\del\tau H_i})_{m_im_{i+1},m'_im'_{i+1}} ,
\end{align}
where $m_i,m'_i$ label of the quantum states of $\v S_i$.  Therefore
the imaginary-time path-integral $\Tr \e^{-\tau H}$
can be expressed as a tensor trace over a tensor network of $T$
(see Fig. \ref{stepY}(a))
\begin{align*}
 \Tr \e^{-\tau H} =\text{tTr} \otimes_i T .
\end{align*}
Now we can use the TEFR approach to evaluate the tensor trace and to
obtain the fixed-point tensors.  This allows us to find the
zero-temperature phase diagram and the quantum phase transitions of
the spin-1 model \eq{spin1}.

\begin{figure}[tb]
\begin{center}
\includegraphics[scale=.5]
{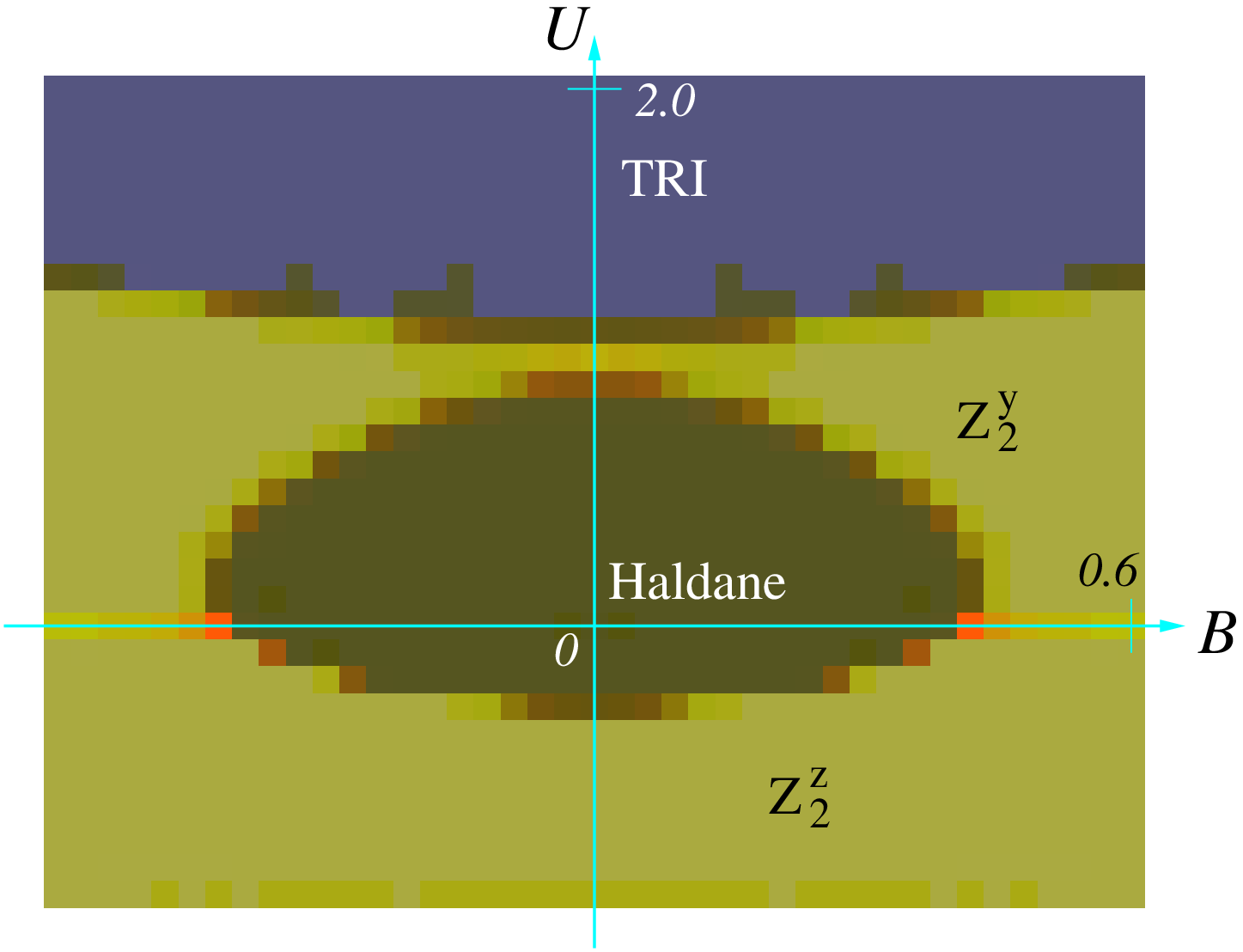}
\end{center}
\caption{
(Color online)
The color-map plot of $(r,g,b)=(X_1/3,X_2/3,1/2\la_\text{sum})$ shows
the phase diagram of our spin-1 system \eq{spin1} which contains four
phases, a trivial spin polarized phase ``TRI'', two $Z_2$ symmetry
breaking phases ``$Z^x_2$'' and ``$Z^y_2$'', and a Haldane phase
``Haldane''.  The transition between those phases are all continuous
phase transitions.
}
\label{X12laSp1xxZ}
\end{figure}
\begin{figure}[tb]
\begin{center}
\includegraphics[scale=.5]
{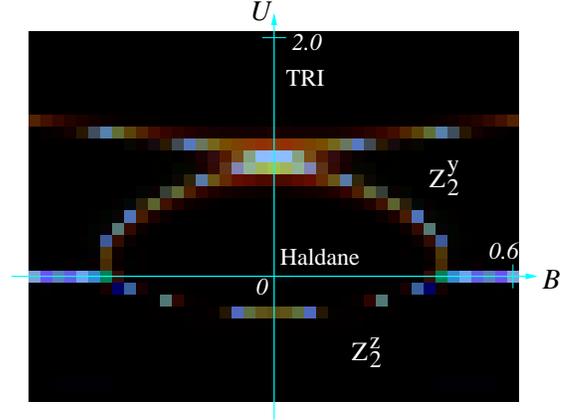}
\end{center}
\caption{
(Color online)
The color-map plot of the central charge as a function of $U,B$.  The
color is given by $(r,g,b)=(c_8,c_9,c_{10})$, where $c_i$ is the central
charge obtained at the $i^{th}$ iteration of Fig. \ref{RG}.  The lines
with non-zero central charge are the critical lines with gapless
excitations.  The model \eq{spin1} has a $U(1)$ spin rotation symmetry
along the $U=0$ line.  The $U=0$ line between the two $Z_2$ phases is
a gapless state.
}
\label{cSp1xxZ}
\end{figure}

However, for small $\del \tau$, the tensor network is very anisotropic
with very different correlation lengths (measured by the distance of
the tensor network) in the time and the space directions.  This causes
large truncation errors in the TEFR steps Fig.
\ref{TEFsqCG}(a)$\to$(b) and Fig. \ref{TEFsqCG}(c)$\to$(d) [or Fig.
\ref{tsrd}(a)].  We need to perform anisotropic coarse graining to
make the system more or less isotropic before performing the TEFR
coarse graining. One such anisotropic coarse graining in time
direction is described in Fig. \ref{stepY}.  Each coarse graining step
of Fig. \ref{stepY} triple the effective $\del\tau$.  After making the
effective $\del\tau$ to be of order 1, we switch to the isotropic
coarse graining in Fig. \ref{TEFsqCG}.

We performed 11 steps of TEFR iteration in Fig.  \ref{RG} (which is
implemented as 22 steps  of TEFR iteration in Fig.  \ref{TEFsqCG}) with
$D_\text{cut}=37$ for various value of $U$ and $B$.  After the TEFR
iteration, a initial tensor $T(U,B)$ flows to a fixed-point tensor. As
we vary $U$ and $B$, the different initial tensors may flow to
different fixed-point tensors which represent different phases.  A
more detailed description of the renormalization flow of the tensors
can be found in Appendix \ref{TEFflow}.

To quantitatively plot the fixed-point tensors, we can use $X_1$,
$X_2$ (see \eq{X12}) , and $\la_\text{sum}$.  Here $\la_\text{sum}$ is
given by $\la_\text{sum} =\sum_s |\la_s/\la_1|$ and $\la_i$ are the
singular values of the matrix $M^{lurd}$ obtained from the fixed-point
tensor (see \eq{TM} and \eq{ULaV}).  In Fig.  \ref{X12laSp1xxZ}, we plot
the $X_1$, $X_2$, and $\la_\text{sum}$ of the resulting fixed-point
tensors for different value of $U$ and $B$.

We see that our spin-1 system has four different phases: TRI, $Z^z_2$,
$Z^y_2$, and Haldane.  The transition between the Haldane phase and
the $Z_2$ phase happens at $B_c=0.405$ along the $U=0$ line, which
agrees with the Haldane gap $\Del =0.4097 = B_c$ obtained in
\Ref{LP8815}.  The transition between the Haldane phase and the TRI
phase happens at $U_c=1.0$ along the $B=0$ line, which agrees with the
result $U_c=0.99$ obtained in \Ref{GJL9254}.

The fixed-point tensor in the TRI phase is the
trivial dimension-one tensor $T^\text{TRI}$ with
$(X_1,X_2,\la_\text{sum})=(1,1,1)$.  So the TRI phase has no symmetry
breaking.  The fixed-point tensor in the $Z^y_2$ and $Z^z_2$ phase,
up to a field redefinition transformation, are the dimension-two tensor
$T^{Z_2}$ which is a direct sum of two dimension-one tensor
$T^\text{TRI}$ with $(X_1,X_2,\la_\text{sum})=(2,2,2)$. So the $Z^y_2$
and the $Z^z_2$ phases are $Z_2$ symmetry breaking phases.
The fixed-point tensor in the Haldane phase
is a dimension-four tensor.  However, many different dimension-four
tensors can appear as the fixed-point tensors. All those
dimension-four fixed-point tensors have
$(X_1,X_2,\la_\text{sum})=(1,1,4)$.  In fact they are all equivalent
to the following CDL tensor
\begin{equation}
\label{THaldane}
T^\text{Haldane}=T(\si^y,\si^y,\si^y,\si^y)
\end{equation}
through the field-redefinition transformation \eq{fred}, where
$T(\si^y,\si^y,\si^y,\si^y)$ is given in \eq{cdlts}.  In fact, one can
show that the following four CDL tensors,
$T(\si^x,\si^x,\si^x,\si^x)$, $T(\si^y,\si^y,\si^y,\si^y)$,
$T(\si^z,\si^z,\si^z,\si^z)$, and $T(\si^0,\si^0,\si^0,\si^0)$, are
all equivalent through the field-redefinition transformation.  (Here
$\si^0$ is the two by two identity matrix.) So they can all be viewed
as the fixed-point tensors in the Haldane phase.  Since the
fixed-point tensor in the Haldane phase is a CDL tensor, the Haldane phase,
like the TRI phase, also does not break any symmetry.

From the fixed-point tensors, we can also obtain the low energy
spectrum of our spin-1 system (see Appendix \ref{partcri}).  Using the
fixed-point tensors of the four phases, we find that all four phases
have a finite energy gap.  We also find that the ground states of the
TRI and Haldane phases are non degenerate while the ground states of
the $Z^y_2$ and $Z^z_2$ phases have a two-fold degeneracy.  This is
consistent with our result that the TRI phase and the Haldane phase do
not break any symmetry while the $Z^y_2$ and $Z^z_2$ phases break a
$Z_2$ symmetry.

The central charge calculated from the resulting fixed-point tensors
is plotted in Fig. \ref{cSp1xxZ} (see \eq{chla}).  The lines of
non-zero central charge in Fig. \ref{cSp1xxZ} mark the region of
diverging correlation length, which correspond to line of continuous
phase transition.  The energy gap closes along those lines.  Compare
Fig.  \ref{cSp1xxZ} with  Fig.  \ref{X12laSp1xxZ}, we see that all the
phase transitions in the phase diagram Fig.  \ref{X12laSp1xxZ}
correspond to those critical lines.  Those phase transitions are all
continuous phase transitions.  The transitions between Haldane and
$Z_2^z$, Haldane and $Z_2^y$, TRI and $Z_2^y$, are described by
central charge $c=1/2$ critical point.  The transitions between
$Z_2^y$ and $Z_2^z$, Haldane and TRI, are described by central charge
$c=1$ critical point.

The continuous phase transition between the TRI and $Z_2$ phases, and
between the Haldane and $Z_2$ phases are $Z_2$ symmetry breaking
transitions.  The phase transition between the $Z^y_2$ and the $Z^z_2$
phases is of the same type as that discussed in section \ref{sp12} along
the $h=0$ line.  Along the $U=0$ line, our spin-1 model has a $U(1)$
spin rotation symmetry around the $S^x$ axis.  The gapless phase at
$B>B_c$ and $U=0$ is close to a $U(1)$ symmetry breaking phase where
spins are mainly in the $y$-$z$ plane with a finite and uniform $S_x$
component.  When $U\neq 0$, the $U(1)$ spin rotation symmetry is
broken. When $U>0$, the spin-1 model
is in the $Z^y_2$ phase with $S^y_i\propto (-)^i$.
When $U<0$, the spin-1 model
is in the $Z^z_2$ phase with $S^z_i\propto (-)^i$.

Both the TRI phase and the Haldane phase have the same symmetry, but
yet they are separated by phase transitions.  This suggests that the
TRI phase and the Haldane phase are distinct phases.  But we cannot
use symmetry breaking to distinguish the two phases.
So let us discuss Haldane phase in more detail.

\subsection{Haldane phase -- a symmetry protected topological phase}

The first question is that weather the TRI phase and the Haldane
phase are really different? (For a detail discuss on the definitions
of phases and phases transitions, see Appendix \ref{phase}.) Can we
find a way to deform the two phases into each without any phase
transition? At first sight, we note that the fixed-point tensor for
the Haldane phase are CDL tensors.  According to the discussion in
section \ref{cdlphy},  this seems to suggest that the Haldane phase is
the trivial TRI phase described by $T^{TRI}$.

In fact, the answer to the above question depend on the symmetry of
the Hamiltonian.  If we allow to deform the spin-1 Hamiltonian
arbitrarily, then the TRI phase and the Haldane phase will belong to
the same phase, since the two phases can be deformed into each other
without phase transition.  If we require the Hamiltonian to have
certain symmetries, then  the TRI phase and the Haldane phase are
different phases (in the sense defined in Appendix \ref{phase}).

To demonstrate the above result more concretely, let us consider a
spin-1 system with time-reversal, parity (spatial reflection), 
and translation symmetries.
The time-reversal symmetry requires the Hamiltonian $H$
and the corresponding tensor $T$ (see \ref{THi}) to be real.  Since
the real Hamiltonian must be symmetric, time-reversal symmetric tensor
must also satisfy
\begin{equation}
\label{timeT}
 T_{ruld}=T_{dlur}
\end{equation}
(Note that the time-reversal transformation exchanges
$(m_i,m_{i+1})\leftrightarrow (m'_i,m'_{i+1})$ in \eq{THi}) The parity
symmetry requires the tensor $T$ to satisfy
\begin{equation}
\label{ParityT}
 T_{ruld}=T_{urdl}.
\end{equation}
(Note that the parity exchanges $(m_i,m'_i)\leftrightarrow
(m_{i+1},m'_{i+1})$ in \eq{THi}) The tensor network considered here is
always uniform which ensure the translation symmetry.  Our spin-1
Hamiltonian \eq{spin1} has time-reversal, parity, and translation
symmetries.

Using our numerical TEFR calculation, we have checked
that the Haldane phase is stable against any perturbations that
respect the time-reversal, parity, and translation symmetries.  In
other words, we start with a tensor $T$ that respects the those
symmetries and flows to the fixed-point tensor $T^\text{Haldane}$.  We
then add an arbitrary perturbation $T\to T+\del T$ that also respect
the time-reversal, parity, and translation symmetries.  We find that
$T+\del T$ also flows to the fixed-point tensor $T^\text{Haldane}$ as
long as $\del T$ is not too large.  This result suggests that  the
time-reversal, parity, and translation symmetries protect the
stability of the Haldane phase. This implies that in the presence of
those symmetries, the TRI phase and the Haldane phase are always
different.  We cannot smoothly deform the Haldane phase into the
TRI phase through the Hamiltonian that have the time-reversal,
parity, and translation symmetries.

If we add a perturbation $\del T$ that has time-reversal symmetry but
not parity symmetry, our numerical result shows that the perturbed
tensor $T+\del T$ will fail to flow to $T^\text{Haldane}$. Instead, it
will flow to the trivial fixed-point tensor $T^\text{TRI}$.  This
implies that, without the parity symmetry, the Haldane phase is
unstable and is the same as the trivial phase TRI, which agrees
the result obtained previously in \Ref{BTG0819}.

To summarize, for Hamiltonian with time-reversal and parity
symmetries, the Haldane phase and the TRI phase are different phases,
despite both phases do not break any symmetries.  So the difference
between the Haldane phase and the TRI phase cannot be describe by
Landau symmetry breaking theory. In this sense the Haldane phase
behaves like the topologically ordered phases.

On the other hand, for Hamiltonian without parity symmetry, the
Haldane phase and the TRI phase are the same phase.  We see that the
Haldane phase exists as a distinct phase only when the Hamiltonian
have time-reversal, parity, and translation symmetries.  This behavior
is very different from the topological phases which are stable against
any perturbations, including those that break the parity symmetry.
Therefore, we should refer Haldane phase as a symmetry protected
topological phase.  We like to stress that the Haldane phase is only
one of many possible symmetry protected topological phases.  Many
other symmetry protected topological phases are studied using
projected symmetry group in \Ref{Wqoslpub,K062}. For example, without
symmetry, there is only one kind of $Z_2$ spin liquid.  In the presence
of spin rotation symmetry and the symmetries of square lattice, there
are hundreds different $Z_2$ spin liquids.\cite{Wqoslpub} All those
$Z_2$ spin liquids can be viewed as symmetry protected topological
phases.

Since the time-reversal, parity, and translation symmetries play such an
important role in the very existence of the Haldane phase, it is not proper to
characterize the Haldane phase just using the fixed-point tensor
$T^\text{Haldane}$. We should use a pair $(G_{TPT}, T^\text{Haldane})$ to
characterize the Haldane phase.  Here $G_{TPT}$ is the symmetry group generated
the time reversal, parity, and translation transformations.  Using the more
complete notation, we can say that $(G_{TPT}, T^\text{Haldane})$ and $(G_{TPT},
T^\text{TRI})$ correspond to different phases, while $(G_{TT},
T^\text{Haldane})$ and $(G_{TT}, T^\text{TRI})$ correspond to the same phase.
Here $G_{TT}$ is the symmetry group generated only by the time-reversal and
translation transformations.  In terms of the new notation, the four phases,
TRI, Haldane, and the two $Z_2$, are labeled by $(G_{TPT}, T^\text{TRI})$,
$(G_{TPT}, T^\text{Haldane})$, $(G_{TPT}, T^{Z_2^y})$.  and $(G_{TPT},
T^{Z_2^z})$.  For a more detailed discussion on how the symmetry
transformations act on the fixed-point tensor, see Appendix \ref{psg}

\begin{figure}[tb]
\begin{center}
\includegraphics[scale=.5]
{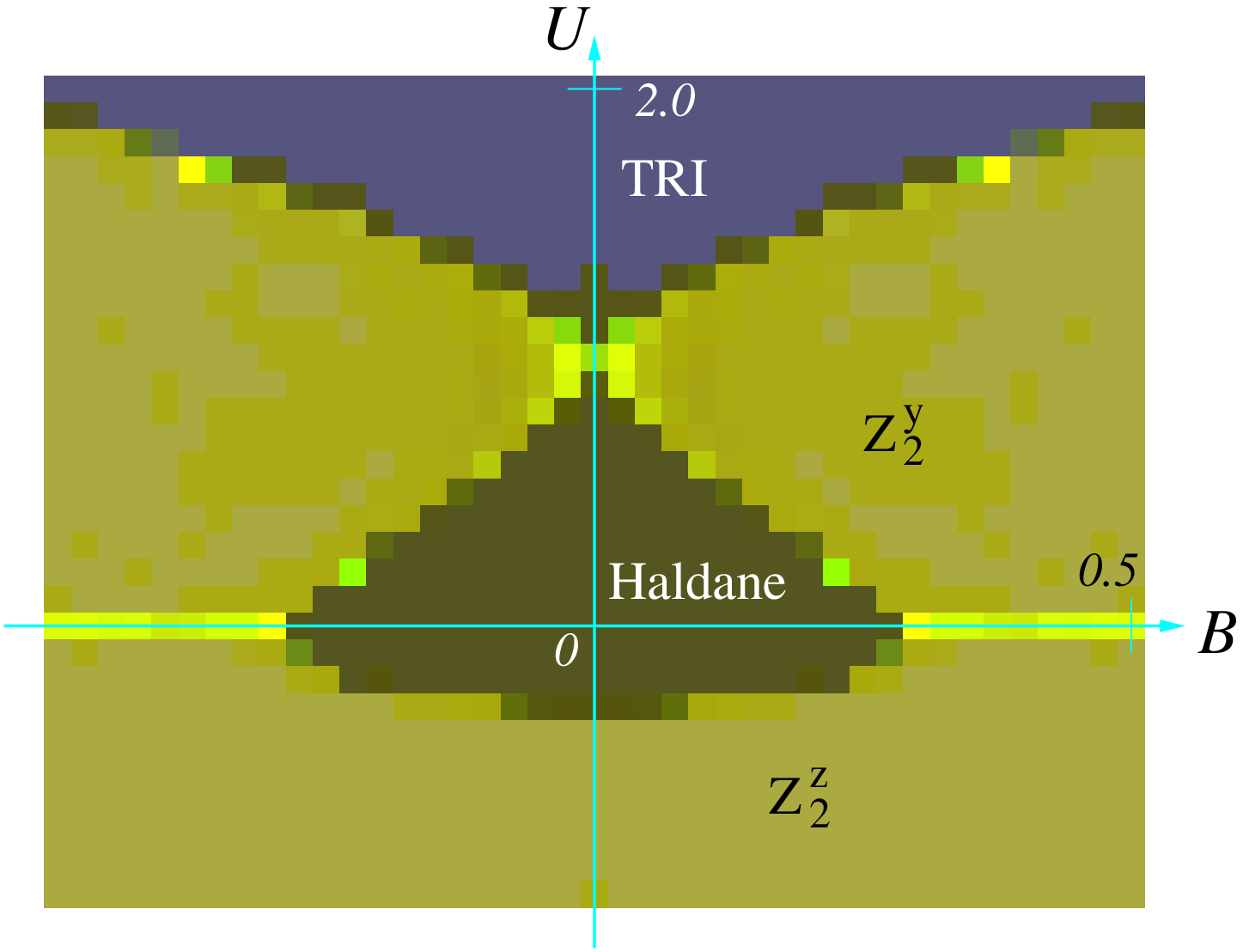}
\end{center}
\caption{
(Color online)
The color-map plot of $(r,g,b)=(X_1,X_2,1/\la_\text{sum})$ shows the
phase diagram of the spin-1 system \eq{spin1XZ} which contains four
phases, a trivial phase ``TRI'', two $Z_2$ symmetry
breaking phases ``$Z^y_2$'' and ``$Z^z_2$'',
and a Haldane phase ``Haldane''.  The
transition between those phases are all continuous phase transitions.
}
\label{X12LASp1}
\end{figure}
\begin{figure}[tb]
\begin{center}
\includegraphics[scale=.5]
{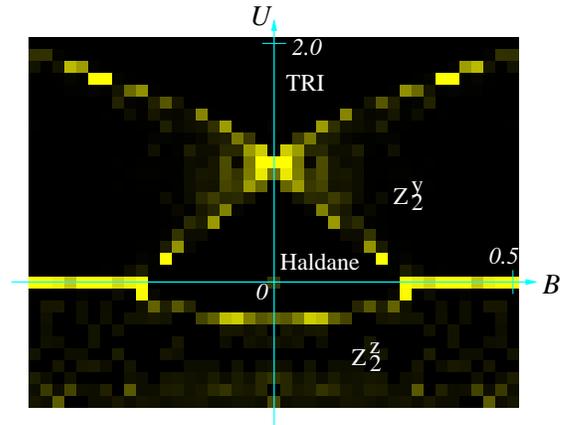}
\end{center}
\caption{
(Color online)
The central charge as a function of $U,B$.  The lines with non-zero
central charge are the critical lines with gapless excitations.  The
model \eq{spin1XZ} has a $U(1)$ spin rotation symmetry along the $U=0$
line.  The $U=0$ line between the two $Z_2$ phases is a gapless state.
}
\label{cSp1}
\end{figure}

In this paper, we propose to use the pair $(G_{TPT},
T^\text{Haldane})$ to characterize the Haldane phase -- a symmetry
protected topological phase.  Such a notation allows us to show that
the Haldane phase is stable against any perturbations that respect
time-reversal, parity, and translation symmetry.  We feel that the new
characterization capture the essence of Haldane phase.  People have
proposed several other ways to characterize the Haldane phase.  One of
them is to use the emergent spin-1/2 boundary spins at the two ends of
a segment of the isotropic spin-1 chain in the Haldane phase.
However, such a characterization fails to capture the essence of
Haldane phase.  The Haldane phase still exists in the presence of an
uniform  magnetic field: $B \sum_i S^z_i$, while the four-fold
degeneracy from the emergent spin-1/2 boundary spins is lifted by such
an uniform  magnetic field.  In the presence of an uniform  magnetic
field, we can no longer use the degeneracy of the boundary spin-1/2
spins to detect the Haldane phase.

People also proposed to use the string order parameter
\begin{align*}
\lim_{|i-j|\to \infty}
\left \< S^z_i e^{\imth \pi \sum_{k=i+1}^{j-1} S^z_k} S^z_j
\right\> \neq 0
\end{align*}
to characterize the Haldane phase.
But the string order parameter vanishes in the presence
of perturbations that contain odd number of spin operators, such as
\begin{align*}
 \del H &= \sum_i A_1 [ (S^z_i)^2 S^z_{i+1} +(S^z_{i+1})^2 S^z_i]
\nonumber\\
&\ \ \ \
+\sum_i A_2 [ (S^x_i)^2 S^x_{i+1} +(S^x_{i+1})^2 S^x_i]
\end{align*}
This is because the total number of $S^z=+1$ and $S^z=-1$ sites mod 2
is not conserved in the presence of the above perturbation.  However,
the vanishing string order parameter does not imply the instability of
Haldane phase.  The above perturbation respects the time-reversal and
parity symmetries.  The Haldane phase is stable against such a
perturbation.  Therefore, the string order parameter also fails to
capture the essence of Haldane phase.

\begin{figure}[tb]
\begin{center}
\includegraphics[scale=.5]
{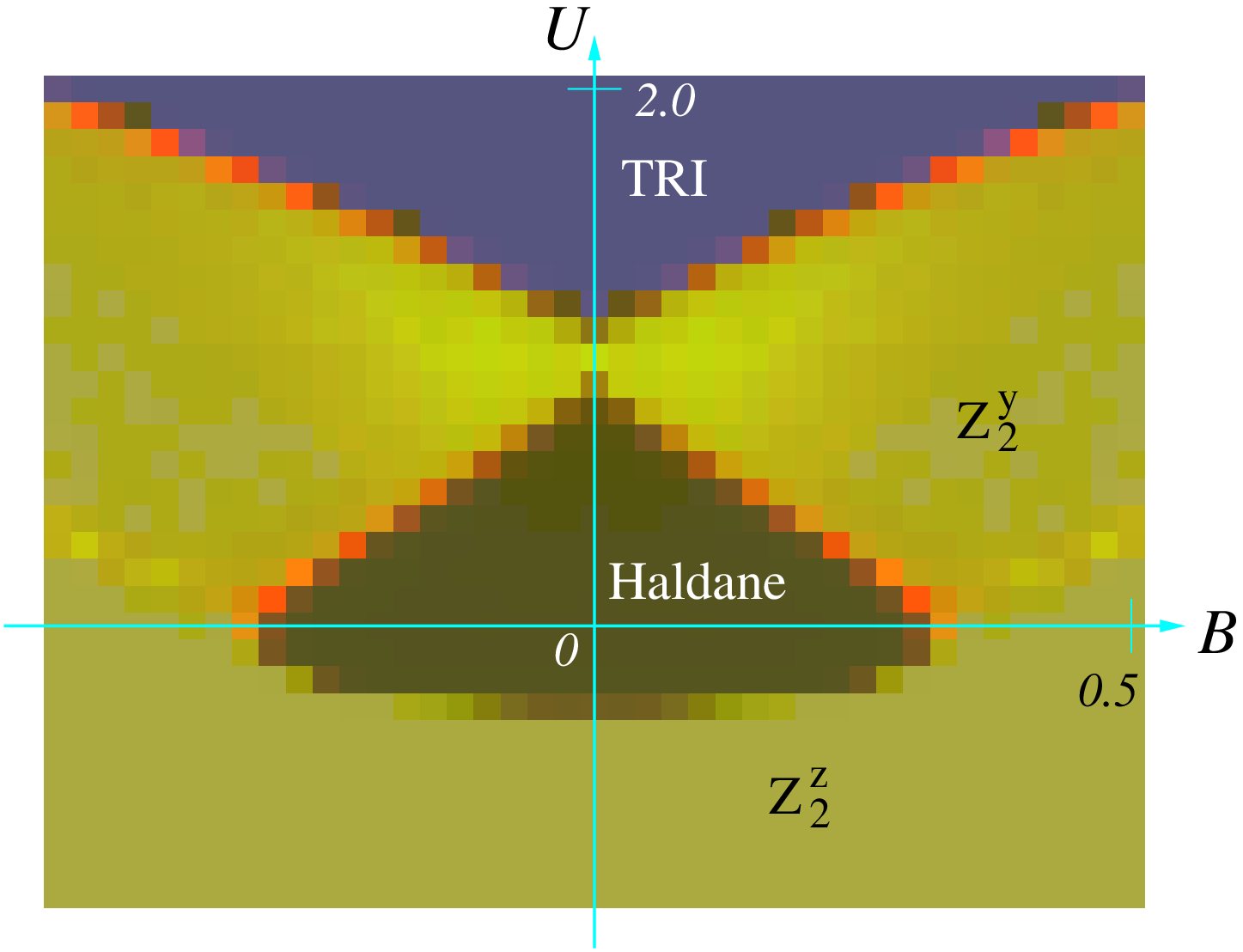}
\end{center}
\caption{
(Color online)
The color-map plot of $(r,g,b)=(X_1/3,X_2/3,1/2\la_\text{sum})$ shows
the phase diagram of the spin-1 system \eq{Sp1xz} which contains four
phases, a trivial spin polarized phase ``TRI'', two $Z_2$ symmetry
breaking phases ``$Z^x_2$'' and ``$Z^y_2$'', and a Haldane phase
``Haldane''.  The transition between those phases are all continuous
phase transitions.
}
\label{X12laSp1xz}
\end{figure}
\begin{figure}[tb]
\begin{center}
\includegraphics[scale=.5]
{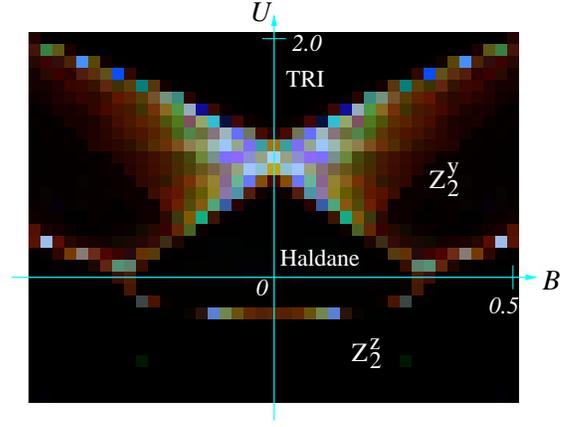}
\end{center}
\caption{
(Color online)
The color-map plot of the central charge as a function of $U,B$.  The
color is given by $(r,g,b)=(c_8,c_9,c_{10})$, where $c_i$ is the central
charge obtained at the $i^{th}$ iteration of Fig. \ref{RG}.  The lines
with non-zero central charge are the critical lines with gapless
excitations.
}
\label{cSp1xz}
\end{figure}

To illustrate the stability of Haldane phase, we also calculated the
phase diagrams of following two spin-1 models.
The first model is given by
\begin{equation}
\label{spin1XZ}
 H=\sum_{i} \left( \v S_{i}\cdot \v S_{i+1} + U (S^z_{i})^2
+ B ( S^x_i +S^z_i)
\right)
\end{equation}
The phase diagram (the color map plot of $X_1$, $X_2$, and
$\la_\text{sum}$) and the central charge are plotted in Figs.
\ref{X12LASp1} and \ref{cSp1}.
The second model is given by
\begin{align}
\label{Sp1xz}
 H & = \sum_{i} \left( \v S_{i}\cdot \v S_{i+1} + U (S^z_{i})^2
\right)
\nonumber\\
&\ \ \
+
\frac{B}{2}
\sum_{i}
( S^x_i(S^z_{i+1})^2 +S^x_{i+1}(S^z_i)^2 + 2S^z_i)
\end{align}
The phase diagram
and the central charge are plotted in Figs.
\ref{X12laSp1xz} and \ref{cSp1xz}.

We see that in both models, the Haldane phase and the TRI phases are
separated by phase transitions, indicating the stability of the
Haldane phase.  (Note that the stability of the Haldane phase is
defined by the impossibility to deform  the Haldane phase to the TRI
phase without a phase transition within a symmetry class of
Hamiltonian.) Both the models have time-reversal and parity symmetries.
In particular, the second model has only time-reversal and parity
symmetries.  This suggests that the  Haldane phase is stable if the
Hamiltonian has time-reversal and parity symmetries.

\begin{figure}[tb]
\begin{center}
\includegraphics[scale=.65]
{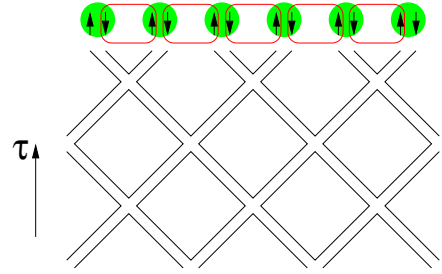}
\end{center}
\caption{
(Color online)
The fixed-point wave function of the Haldane phase can be obtained
from the imaginary-time evolution of the fixe-point tensor network of
the Haldane phase $T^\text{Haldane}=T(\si^y,\si^y,\si^y,\si^y)$, which
is formed by the corner matrix $\si^y$. (\ie the boundary of the
tensor network gives raise to the fixed-point wave function.)  The
dimension two of the corner matrix implies that each index of the
corner matrix is associated with a pseudo spin-1/2.  Thus each coarse
grained site has two spin-1/2 pseudo spins.  The corner matrix $\si^y$
ties the two pseudo spins on the neighboring sites into a pseudo spin
singlets.  Thus the fixed-point wave function of the Haldane phase can
be viewed as a dimer state formed by nearest neighbor pseudo spins.
}
\label{cdlstate}
\end{figure}

Before ending this section, we like to point out that the form of the
fixe-point tensor of the Haldane phase
$T^\text{Haldane}=T(\si^y,\si^y,\si^y,\si^y)$ implies the following
fixed-point wave function for the Haldane phase.  Since the
fixed-point tensor $T^\text{Haldane}=T(\si^y,\si^y,\si^y,\si^y)$ has a
dimension four, the effective degrees of freedom in the fixed-point
wave function is four states per site (\ie per coarse grained site).
The four states can be viewed as two spin-1/2 pseudo spins.  The
fixed-point wave function is then formed by the pseudo spin singlet
dimers between nearest neighbor bonds (see Fig. \ref{cdlstate}).

\section{Summary}

\begin{figure}[tb]
\begin{center}
\includegraphics[scale=.49]
{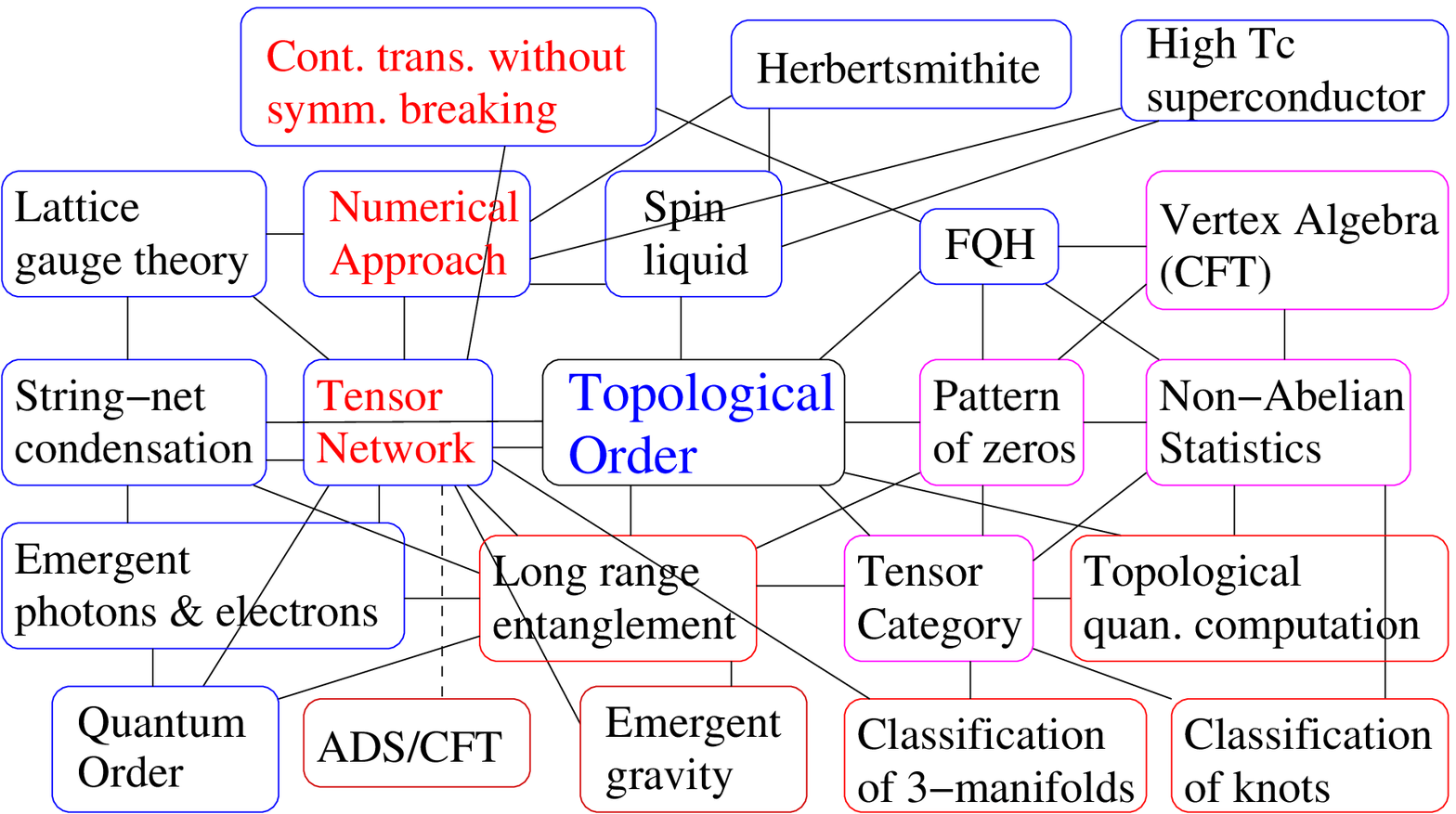}
\end{center}
\caption{
(Color online)
The paradigm of topological order.
}
\label{TopPara}
\end{figure}

In this paper, we introduce a TEFR approach for statistical and quantum
systems.  We can use the TEFR approach to calculate phase diagram with
both symmetry breaking and topological phases.  We can also use the
TEFR approach to calculate critical properties of continuous
transitions between symmetry breaking and/or topological phases.  The
TEFR approach suggests a very general characterization of phases based on
the symmetry group $G_\text{sym}$ and the fixed-point tensor
$T_\text{inv}$. The $(G_\text{sym},T_\text{inv})$ characterization can
describe both symmetry breaking and topological phases.  Using such a
characterization and through the stability of the fixed-point tensor,
we show that the Haldane phase for spin-1 chain is stable against any
perturbations that respect time-reversal, parity, and translation
symmetry. This suggests that the Haldane phase is a symmetry protected
topological phase.

The TEFR approach introduced here can be applied to any strongly
correlated systems in any dimensions.  It has a potential to be used
as a foundation to study topological phases and symmetry protected
topological phases in any dimensions. In this paper, we only
demonstrate some basic applications of the TEFR approach through some
simple examples. We hope those results will lay the foundation for
using the TEFR approach to study more complicated strongly correlated
systems in 1+2 and in 1+3 dimensions.

To gain a better understanding of the tensor network renormalization
approach, let us compare the tensor network renormalization approach
with the block-spin renormalization approach.\cite{K6663} Let us
consider a spin system whose partition function can be represented by
a tensor network of a tensor $T(g)$, where $g$ is a coupling constant
of the spin system.  After performing a coarse graining of the
lattice, the system is described by a coarse grained tensor network of
tensor $T'$.  The transformation $T(g)\to T'$ represent tensor network
renormalization flow.  In the block-spin renormalization approach, we
insist on the coarse grained system have the same form as described by
a new coupling constant $g'$.  So we try to find a $g'$ such that
$T(g')$ best approximates $T'$.  This way, we obtain a block-spin
renormalization flow $g\to g'$.  We see that tensor network
renormalization flow is a transformation in the generic tensor space,
while the block-spin renormalization flow is a transformation is the
subspace $T(g)$ parameterized by the coupling constant $g$.  We believe
that it is this generic flow in the whole tensor space allows the
tensor network renormalization approach to describe topological
phases.

In this paper, we stress that the fixed-point tensor of the tensor
network renormalization flow can be used as a quantitative description
of topological order.  This is just one aspect of topological order.
In fact, theory of topological order has a very rich structure and
covers a wide area (see Fig. \ref{TopPara}).  Turaev-Viro
invariant\cite{TV9265} on 3D manifold can be viewed as a tensor trace
of the fixed-point tensors for string-net models.\cite{LWstrnet}
Therefore, the tensor network renormalization approach should apply to
all time reversal and parity symmetric topological orders in 1+2
dimensions which are described by the string-net
condensations.\cite{FNS0428,LWstrnet} This type of topological order
include emergent gauge theory, Chern-Simons theory, and quantum
gravity in 1+2 dimensions.\cite{MT9295,S9503,LWstrnet} The
mathematical foundation that classifies this type of fixed-point
tensors is the tensor category theory. It is natural to expect that
the topological order in 1+3 dimensions and the associated emergent
gauge bosons, emergent fermions, and emergent
gravitons\cite{Wqoem,GW0600} can also be studied through the tensor network
renormalization approach.  The tensor network renormalization approach
also provides a new way to calculate lattice gauge theories.  We see
that physics is well inter connected.  It appears that the tensor
network renormalization is a glue that connects all different parts of
physics.

\section{Acknowledgements}
We would like to thank Michael Levin, Frank Verstraete, and Michael
Freedman for many very helpful discussions.  This research was
supported by the Foundational Questions Institute (FQXi) and NSF Grant
DMR-0706078.

\appendix

\section*{Appendix}

\subsection{A detailed discussion of the TEFR algorithm}
\label{teftnr}

The key step in the TEFR calculation is the entanglement filtering
operation from Fig. \ref{TEFsqCG}(b) to Fig. \ref{TEFsqCG}(c).
In this section we will discuss several ways to implement such
an entanglement filtering calculation.

\subsubsection{The linear algorithm}
To find those new $S^\prime_i$ described in Fig. \ref{S1234}, we may
minimize the following cost function:
\begin{align}
\label{LC}
& \ \ \ C_L=
\\
&
\parallel
\sum_{ijkl}{S_1}_{li\alpha}{S_2}_{ij\beta}
{S_3}_{jk\gamma}{S_4}_{kl\delta}
-\sum_{ijkl}{S_1}^\prime_{li\alpha}{S_2}^\prime_{ij\beta}
{S_3}^\prime_{jk\gamma}{S_4}^\prime_{kl\delta}
\parallel
\nonumber
\end{align}
where $\alpha,\beta,\delta,\gamma=1,\ldots D$ and $i,j,k,l=1,\ldots
D^\prime$

This cost function can be minimized by solving a sets of least
square problems iteratively. For example, we may first fixed
$S^\prime_2,S^\prime_3,S^\prime_4$ and solve a least square problem
for $S_1^\prime$, then we fixed $S^\prime_1,S^\prime_3,S^\prime_4$
to solve a least square problem for $S_2^\prime$, etc. In practise,
to reduce the trapping by local minima, we may start with
$D^\prime=D-1$ and gradually decrease $D^\prime$.

\subsubsection{The nonlinear algorithm}

Although the linear algorithm is an efficient way to do the
entanglement-filtering calculation in Fig. \ref{S1234}, however, this
algorithm could not avoid trapping by local minima in generic case. To
solve this problem, here we introduce a non-linear algorithm which can
avoid the trapping by local minima.

In the first implementation of TEFR discussed above, we choose $S'_i$
to minimize the dimension of the one of the index of the rank-three
tensors $S'_i$.  The second implementation of TEFR
is still described by Fig. \ref{TEFsqCG}.  But now we choose $S'_i$
to minimize the entanglement entropy on the diagonal links in Fig.
\ref{TEFsqCG}(d).

\begin{figure}[tb]
\begin{center}
\includegraphics[scale=0.8]
{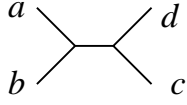}
\end{center}
\caption{
A rank-four tensor expressed as tensor network.
}
\label{YY}
\end{figure}

Let us define the entanglement entropy on a link.  We may view the
tensor network in Fig. \ref{YY}, as matrix $M_{i,j}$ where $i=(a,b)$
and $j=(c,d)$. We then perform an SVD of $M$ which produces the
singular values $\la_i$ that are ordered as $\la_1\geq \la_2 \geq ...
\geq 0$. We call $\la_i$ the singular values on the horizontal link in
Fig. \ref{YY}.  The entanglement entropy on the horizontal link in
Fig. \ref{YY} is defined as $S=-\sum_i \t\la_i \log (\t\la_i)$, where
$\t\la_i=\la_i/\sum \la_i$.  In the  non-linear algorithm, we choose $
S'_1$, $S'_2$, $S'_3$, and $S'_4$ to minimize the following cost
function
\begin{align}
\label{NLC}
& \ \ \ C_{NL}= \lambda S_\text{sum}+
\\
&
\parallel
\sum_{ijkl}{S_1}_{li\alpha}{S_2}_{ij\beta}
{S_3}_{jk\gamma}{S_4}_{kl\delta}
-\sum_{ijkl}{S_1}^\prime_{li\alpha}{S_2}^\prime_{ij\beta}
{S_3}^\prime_{jk\gamma}{S_4}^\prime_{kl\delta}
\parallel
\nonumber
\end{align}
where $S_\text{sum}$ is the sum of entanglement entropy on the four
diagonal links in Fig.  \ref{TEFsqCG}(d).  Here $\lambda$ should be as
small as possible.

We note that when we reduce the entanglement entropy on the diagonal
links in Fig. \ref{TEFsqCG}(d), the singular values on those diagonal
links decay faster. This allows us to drop indices associated with the
very small singular values and hence reduce the dimension on those
diagonal links. When we use Fig. \ref{tsrd}(b) to transform Fig.
\ref{TEFsqCG}(d) to Fig. \ref{TEFsqCG}(e), the resulting tensor $T'$ will
have a small dimension. This approach can again reduce a CDL tensor
$T$ to a trivial dimension-one tensor.

The non-linear algorithm is a very general approach, but the
calculation is very expensive. In the paper, we will not use this
algorithm.

\subsubsection{The scaling-SVD algorithm for CDL tensors}

In this section, we introduce the third algorithm, which is very simple and
efficient.  We will call such algorithm scaling-SVD algorithm, which
can be used to find $S'_1$, $S'_2$, $S'_3$, and $S'_4$ in Fig.
\ref{S1234} with smaller dimensions.

\begin{figure}[tb]
\begin{center}
\includegraphics[scale=0.7]
{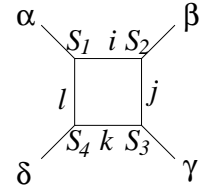}
\end{center}
\caption{The entanglement-filtering procedure. } \label{clean}
\end{figure}

In the scaling-SVD algorithm,
we introduce four diagonal weighting matrices:
\begin{align*}
w^{12}_{i i^\prime}&=w_i^{12}\delta_{ii^\prime};\qquad w^{23}_{j
j^\prime}=w_j^{23}\delta_{jj^\prime}; \nonumber\\w^{34}_{k
k^\prime}&=w_k^{34}\delta_{kk^\prime};\qquad w^{41}_{l
l^\prime}=w_l^{41}\delta_{l l^\prime};
\end{align*}
on those links $i,j,k,l$ in Fig. \ref{clean} and initialize them as
$1$.  We first perform the scaling-SVD calculation on the pair $S_1$
and $S_2$ by introducing a matrix:
\begin{equation*}
M^{12}_{\alpha l;j\beta}=\sum_i
\sqrt{w_l^{41}}{S_1}_{li\alpha}{S_2}_{ij\beta}\sqrt{w_j^{23}}
\end{equation*}
and do the SVD decomposition $M^{12}=U\Lambda V^\dagger$:
\begin{equation*}
M^{12}_{\alpha l;j\beta}=\sum_i U_{\alpha
l;i}\Lambda_i^{12}V_{j\beta;i} .
\end{equation*}
We then
update $S_1,S_2$ as:
\begin{equation*}
{S_1}^\prime_{li\alpha}=\sqrt{\Lambda^{12}_i}U_{\alpha
l;i}/\sqrt{w_l^{41}}; \qquad
{S_2}^\prime_{ij\beta}=\sqrt{\Lambda^{12}_i}V_{j\beta;i}/\sqrt{w_j^{23}}
\end{equation*}
We can update $S_2,S_3$ in the same way, then we can further update
$S_3,S_4$ and $S_4,S_1$.
After one loop update for those $S_i$'s, we update all the weighting
factors $w^{12}$ \etc as $w^{12}=\Lambda^{12}$ \etc on corresponding links.

After several loop iterations, we truncate the smallest singular
values of
$\Lambda^{12}_i,\Lambda^{23}_j,\Lambda^{34}_k,\Lambda^{41}_l$ and
obtain new rank-three tensors
$S_1^\prime,S_2^\prime,S_3^\prime,S_4^\prime$ with a smaller dimension
$D^\prime<D$ on the inner links.
If $S_1$, $S_2$, $S_3$, and $S_4$ are CDL tensor, we find:
\begin{equation*}
{S_1}^\prime_{li\alpha}{S_2}^\prime_{ij\beta}
{S_3}^\prime_{jk\gamma}{S_4}^\prime_{kl\delta} \propto
{S_1}_{li\alpha}{S_2}_{ij\beta} {S_3}_{jk\gamma}{S_4}_{kl\delta}
\end{equation*}
after truncation. By proper normalizing those $S^\prime_i $, we can
find the solution of
\begin{equation}
\label{SSp}
\sum_{ijkl}{S_1}_{li\alpha}{S_2}_{ij\beta}
{S_3}_{jk\gamma}{S_4}_{kl\delta}
-\sum_{ijkl}{S_1}^\prime_{li\alpha}{S_2}^\prime_{ij\beta}
{S_3}^\prime_{jk\gamma}{S_4}^\prime_{kl\delta} \approx 0 .
\end{equation}
If $S_1$, $S_2$, $S_3$, and $S_4$ are not CDL tensor, the truncated
$S'_1$, $S'_2$, $S'_3$, and $S'_4$ cannot satisfy \eq{SSp} even after
the rescaling of $S'_1$, $S'_2$, $S'_3$, and $S'_4$.  In this case, we
reject the truncated $S'_1$, $S'_2$, $S'_3$, and $S'_4$ and the
scaling-SVD fails to produce simplified tensors.  To summarize, the
scaling-SVD algorithm can simplify CDL tensors, but not other tensors.

To understand why the above scaling-SVD algorithm can work for CDL
tensors, we can consider the simplest CDL tensor
$T_{ruld}(M^1,M^2,M^3,M^4)$ with corner matrices
\begin{equation*}
M^1=M^2=M^3=M^4=\bpm
1 & 0  \\
0 & x  \\
\epm .
\end{equation*}
where $0<x<1$.

\begin{figure}[tb]
\begin{center}
\includegraphics[scale=1.5]
{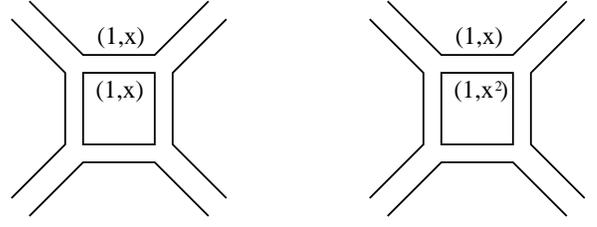}
\end{center}
\caption{For the simplest CDL tensor, the outer line singular values
are not scaled after one loop iteration, however, the inner line
singular values are scaled to a power. This nontrivial scaling
behavior is the key why the scaling-SVD algorithm can filter out
those local entanglement for a CDL tensor.} \label{cdlRG}
\end{figure}

See in Fig. \ref{cdlRG}, after one loop iteration, the
$4\times4$ diagonal weighting matrix $w=w_{In}\otimes w_{Out}$ on
each link is updated as:
\begin{equation*}
w_{In}=\bpm
1 & 0  \\
0 & x  \\
\epm; \qquad w_{Out}=\bpm
1 & 0  \\
0 & x  \\
\epm
\end{equation*}
Where $In$ represents the inner line and $Out$ represent the outer
line of the square as show in Fig. \ref{cdlRG}. However, after
$n$ loop iteration, we find:
\begin{equation*}
w_{In}=\bpm
1 & 0  \\
0 & x^n  \\
\epm; \qquad w_{Out}=\bpm
1 & 0  \\
0 & x  \\
\epm
\end{equation*}
The striking thing is that the weighting matrices on the outer lines
are never scaled but the weighting matrices (as well as the singular
values on the corresponding link) on the inner lines are scaled to its
$n$ power! After enough times of loop iterations, $x^n$ can be smaller
than the machine error.  The inner line can be removed and we
successfully filter out those local entanglements in the square.  In
this paper, we do all our TEFR calculations using this scaling-SVD
algorithm.

It also clear that if the corner matrices in a CDL tensor have
degenerate eigenvalues, the scaling-SVD cannot simplify such a CDL
tensor. This is a desired result, since the degeneracy in the singular
values may indecate the presence of topological order (see the
discussion at the end of Appendix \ref{TEFflow}).

\subsection{Partition function and critical properties
from the tensor network renormalization flow of the tensors}
\label{partcri}

In this section, we will give a more detailed discussion about the
renormalization flow of the tensor.  The more detailed discussion
allows us to see the relation between the fixed-point tensor and the
energy spectrum for a quantum system.  It also allows us to see the
relation between the fixed-point tensor at the critical point and the
scaling dimensions and central charge of the critical point.  To be
concrete, we will consider the renormalization transformation
described by Fig. \ref{RG}, which represents a single iteration step.
Such an iteration step roughly correspond to two iteration steps
described by Fig.  \ref{tsrd}.

We start with a tensor network formed by $N_0=N$ tensors
$T^{(0)}=T$.  After $i$ iterations, the tensor network is deformed into
a tensor network of $N_i$ $T^{(i)}$-tensors.
We formally denote the tensor network renormalization flow as
\begin{equation*}
 T^{(i)} \to T^{(i+1)}.
\end{equation*}
Those tensors satisfy the following defining property
\begin{equation*}
  \text{tTr} (\otimes T^{(i)})^{N_i}
= \text{tTr} (\otimes T^{(i+1)})^{N_{i+1}}  .
\end{equation*}

The scales of the tensors $T^{(i)}$ changes under each iteration.
It is more convenient to describe the tensor network renormalization
 flow in terms of the
normalized tensor.  The normalized tensor is defined as
\begin{equation}
\label{TcT}
\cT^{(i)}=T^{(i)}/\Ga_i
\end{equation}
that satisfies
\begin{equation*}
 \sum_{l,u} \cT^{(i)}_{l,l,u,u}=1
\end{equation*}
In terms of the normalized tensor, after a tensor network
renormalization
transformation, a tensor network of $N_i$ normalized $\cT^{(i)}$-tensors
is transformed into a tensor network of $N_{i+1}$ normalized
$\cT^{(i+1)}$-tensors:
\begin{equation*}
 \cT^{(i)} \to f_{i+1} \cT^{(i+1)} .
\end{equation*}
The $N_i$ normalized $\cT^{(i)}$-tensors and the $N_{i+1}$
$f_i\cT^{(i+1)}$-tensors have the same tensor trace
\begin{equation}
\label{cTcT}
                    \text{tTr} (\otimes \cT^{(i)})^{N_i}
= f_{i+1}^{N_{i+1}} \text{tTr} (\otimes \cT^{(i+1)})^{N_{i+1}} .
\end{equation}
$f_i$ is a quantity that can be directly obtained from the numerical
tensor network renormalization calculation.

From \eq{TcT} we see that the partition function at $i^\text{th}$
iteration can be expressed as a tensor trace of the normalized tensor
$\cT^{(i)}$ up to an over all factor
\begin{align}
\label{ZcT}
 Z_i= \Ga_i^{N_i} \text{tTr} (\otimes \cT^{(i)})^{N_i}.
\end{align}
Note that by definition, the partition function
is invariant under the tensor network renormalization transformation $ Z_i=Z_{i+1}$.
From \eq{cTcT}, we find that $\Ga_i$ satisfies
\begin{align}
\label{GaGa}
\Ga_i^{N_i} f_{i+1}^{N_{i+1}}= \Ga_{i+1}^{N_{i+1}}
\end{align}
which allows us to calculate $\Ga_i$ from $f_i$.  We can keep
performing the tensor network renormalization until $N_i$ is of order 1.  This
way, we can calculate the partition function $Z=Z_i$ from \eq{ZcT} and
\eq{GaGa}.

\begin{figure}
\begin{center}
\includegraphics[scale=0.5]{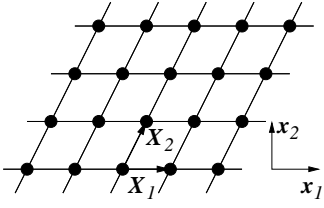}
\end{center}
\caption{
A square-lattice tensor network in a 2D space
(or 1+1D space time).
}
\label{tsnkSp}
\end{figure}

To calculate the density of the log partition function $\log (Z)$, we
need to describe how tensor positions in the tensor network are mapped
into the physical coordinates $\v r=(x_1,x_2)$.  Let $\v X_1$ be the
vector in physical space connecting the two neighboring tensors in the
horizontal direction and $\v X_2$ the vector connecting the two
neighboring tensors in the vertical direction (see Fig. \ref{tsnkSp}).
Let denote the coordinates of $\v X_i$ as
\begin{equation}
\label{X12W}
 \v X_\al=
\bpm W_{1\al} \\ W_{2\al} \epm , \ \ \ \ \al=1,2 .
\end{equation}
The two by two matrix $W$ describes the shape of the tensor network.

We note that the tensor network is changed under each iteration.  So
$W_{\al\bt}$ is also changed under each iteration.  After the $i^{th}$
iteration, the shape of the resulting tensor network is described by
$W^{(i)}_{\al\bt}$, where $W^{(0)}_{\al\bt}=W_{\al\bt}$ is for the
initial tensor network.  Under the renormalization transformation
described in Fig. \ref{RG}, $W^{(i)}$ transforms in a simple way
\begin{equation*}
 W^{(i+1)}= 2W^{(i)}.
\end{equation*}
which implies $A_{i+1}=4A_i$.  Under the renormalization
transformation described in Fig. \ref{CGsq}, $W^{(i)}$ transforms as
\begin{equation*}
 W^{(i+1)}= W^{(i)}
\bpm
1 & 1 \\
-1 & 1 \\
\epm
.
\end{equation*}
The above recursion relations allow us to calculate $W^{(i)}$ for each
step of iteration. Such information will be useful later.

We see that the area occupied by each tensor $T^{(i)}$
after the $i^{th}$ iteration is
\begin{equation*}
 A_i=\text{det} (W^{(i)})
\end{equation*}
where $W^{(i)}$ is the two by two matrix formed by $W^{(i)}_{\al\bt}$.
It is also clear that
\begin{equation*}
A_iN_i =A_{i+1}N_{i+1} =A_\text{tot}
\end{equation*}
is the total area of the system.

Let us introduce
\begin{equation}
z_i=\Ga_i^{1/A_i}  .
\end{equation}
From \eq{GaGa}, we see that $z_i$ satisfies the following
recursion relation
\begin{align}
\label{zzf}
 z_{i+1}=z_{i} f_{i+1}^{1/A_{i+1}}.
\end{align}
or
\begin{equation}
\label{logz}
 \log( z_i) = \log(z_0) + \sum_{j=1}^{i} \frac{ \log(f_{j}) }{A_j}.
\end{equation}
which will allows us to calculate the density of the
log partition function
\begin{equation*}
A_\text{tot}^{-1}\log (Z_i)= \log (z_i)
+ A_\text{tot}^{-1} \log \left(\text{tTr} (\otimes \cT^{(i)})^{N_i} \right) .
\end{equation*}
Note that, after a number of iterations, we can reduce $N_i$ to be of
order 1. In this case $\log (z_i)$ itself is the density of the log
partition function in the large $A_\text{tot}$ limit.

In the large iteration limit, $\cT^{(i)}$ and $f_i$ become independent
of $i$.  This allows us to introduce scale invariant tensor
\begin{equation*}
 T^{(i)}_\text{inv} =\ga^{-1} \cT^{(i)}
\end{equation*}
that satisfy
\begin{equation*}
 T^{(i)}_\text{inv}  \to  T^{(i+1)}_\text{inv}
\end{equation*}
in large $i$ limit.
We note that the invariant tensor has the following defining property
\begin{equation*}
  \text{tTr} (\otimes \cT^{(i)}_\text{inv})^{N_i}
= \text{tTr} (\otimes \cT^{(i+1)}_\text{inv})^{N_{i+1}}
\end{equation*}
From \eq{cTcT}, we see that $\ga$ must satisfy $
\ga^{N_i}=\ga^{N_{i+1}} f_{i+1}^{N_{i+1}}$ or
\begin{align}
\ga= f_i^{1/3} ,
\end{align}
where we have used
\begin{equation}
\label{NiNi1}
N_i=4N_{i+1}.
\end{equation}
Thus the invariant tensor is simply given by
\begin{equation}
\label{Tinv}
 T^{(i)}_\text{inv} =f_i^{-1/3} \cT^{(i)}
\end{equation}

In the large $i$ limit, $f_i$ approaches a constant and the recursion
relation \eq{zzf} implies that
\begin{align*}
& \log (z_{i+1} f_{i+1}^{\frac13\frac{1}{A_{i+1}}}) - \log (z_i f_i^{\frac13\frac{1}{A_i}} )
\nonumber\\
& = \frac{1}{3A_i}  (\log (f_{i+1}) - \log (f_i) )=o(1/A_i) .
\end{align*}
This result allows us to calculate the finite-size correction to the
partition function.  For a tensor network of $N$ tensors $T$, from the
relation $Z=\text{tTr} (\otimes \cT)^{N} = \Ga_i^{N_i} \text{tTr}
(\otimes \cT^{(i)})^{N_i}$, we find that the partition function is
given by
\begin{align*}
 Z= z_i^{A_\text{tot}} \text{tTr} (\otimes \cT^{(i)})^{N_i}
=(z_i f_i^{\frac13\frac{1}{A_i}})^{A_\text{tot}} \text{tTr}  (\otimes \cT^{(i)}_\text{inv})^{N_i}
\end{align*}
When $A_\text{tot}$ and $A_i$ are of the same order (\ie when $N_i$ is
of order 1), we find that
\begin{equation*}
 \log( Z)= A_\text{tot}\log  (z_\infty) +
\log \Big(\text{tTr}  (\otimes T^{(i)}_\text{inv})^{N_i} \Big)
+o(1) .
\end{equation*}
where $z_\infty=\lim_{i\to\infty} z_i f_i^{\frac{1}{3 A_i}}$.  Also note that
the invariant tensor $T^{(i)}_\text{inv}$ is independent of $i$ in
the large $i$ limit (because $T^{(i)}_\text{inv}$ approaches to a
fixed-point tensor).  We see that $\log (Z)$ contains a term
$A_\text{tot}\log  (z_\infty)$ that is proportional to the system area
$A_\text{tot}$.  $\log (Z)$ also contains a $A_\text{tot}$ independent
term $\log \Big(\text{tTr} (\otimes T^{(i)}_\text{inv})^{N_i} \Big)$.  The
remaining $o(1)$ term approaches to zero as $A_\text{tot}\to \infty$.

It is interesting to note that although the  $\log
\Big(\text{tTr} (\otimes T_\text{inv})^{N_i} \Big)$ term in the
partition function does not depend on the total area $A_\text{tot}$ of
the tensor network, it can depend on the shape of the network if the
system is at the critical point.  We can use this shape-dependent
partition function to obtain the critical properties of the critical
point.  In the following, we will study the shape dependence of the
partition function at a critical point.  We will assume that in the
1+1D space time, the velocity is equal to 1 (or the critical system is
isotropic in the 2D space).

One way to obtain shape-dependent partition function (\ie the
$A_\text{tot}^0$ part of the partition function) is to introduce four
matrices associated with the invariant fixed-point tensor
$T^{(i)}_\text{inv}$
\begin{align}
\label{TM}
& (M^{ud})_{u,d}=\sum_r (T^{(i)}_\text{inv})_{rurd},
&&
 (M^{lr})_{l,r}=\sum_u (T^{(i)}_\text{inv})_{rulu},
\nonumber\\
& (M^{ldru})_{ld,ru}= (T^{(i)}_\text{inv})_{ruld},
&&
 (M^{lurd})_{lu,rd}= (T^{(i)}_\text{inv})_{ruld}.
\end{align}
The eigenvalues of the those matrices encode the information about
that shape-dependent partition function.

To understand the physical meaning of those eigenvalues, let us
consider a $T^{(i)}_\text{inv}$-tensor network that has a rectangular
shape $N_i=L\times M$. The shape dependent part of the partition
function is given by $ \text{tTr} (\otimes T^{(i)}_\text{inv})^{N_i}$.
To describe the shape quantitatively, we note that the
$T^{(i)}_\text{inv}$-tensor network corresponds to a parallelogram
spanned by two vectors $\v l=l_\al\v x_\al=L W^{(i)}_{1\al}\v x_\al$ and
$\v m=m_\al\v x_\al=M W^{(i)}_{2\al}\v x_\al$, where $\v x_\al$,
$\al=1,2$ form an orthonormal basis of the space.  The shape, by
definition, is described by a complex number
\begin{align*}
 \tau^{(i)}=\frac{m_1+\imth m_2}{l_1+\imth l_2}
=\frac{W^{(i)}_{21}+\imth W^{(i)}_{22} }{W^{(i)}_{11}
+\imth W^{(i)}_{12}}\frac{M}{L} .
\end{align*}
Many critical properties of the critical point can be calculated from
the $\tau$ dependence of $ \text{tTr} (\otimes
T^{(i)}_\text{inv})^{N_i}$.  In particular (according to conformal
field theory)\cite{C8686}
\begin{align}
\label{Zch}
&\ \ \
\Big(\text{tTr} (\otimes T^{(i)}_\text{inv})^{N_i} \Big)
\nonumber\\
&=\sum_{n=0} e^{-2\pi
\left[
(h^R_n+h^L_n-\frac{c}{12})\Im \tau^{(i)}
+\imth(h^R_n-h^L_n)\Re \tau^{(i)} \right]
} ,
\end{align}
where $c$ is the central charge and $\{h^R_n,h^L_n\}$ is the spectrum
of right and left scaling dimensions of the  critical point.

Since $T^{(i)}_\text{inv}$ is already a fixed-point tensor,
we may evaluate \eq{Zch} by choosing $L=1$:
\begin{align*}
&\ \ \ \text{tTr} (\otimes T^{(i)}_\text{inv})^M =\text{Tr} (M^{ud})^M
\nonumber\\
&=\sum_{n=0} e^{-2\pi M
\left[
(h^R_n+h^L_n-\frac{c}{12})\Im \tau^{(i)}_0
+\imth(h^R_n-h^L_n)\Re \tau^{(i)}_0 \right]
} .
\end{align*}
where
\begin{align}
 \tau^{(i)}_0
=\frac{W^{(i)}_{21}+\imth W^{(i)}_{22} }{W^{(i)}_{11}+\imth W^{(i)}_{12}} .
\end{align}
Using the above result, and some other results obtained with similar
methods, we find that
\begin{widetext}
\begin{align}
\label{chla}
(h^R_n+h^L_n-\frac{c}{12})\Im (\tau^{(i)}_0 )
+\imth(h^R_n-h^L_n)\Re (\tau^{(i)}_0 )
&=-\frac{\log (\la^{ud}_n)}{2\pi} +\text{ mod } \imth,
\nonumber\\
 (h^R_n+h^L_n-\frac{c}{12})\Im (\frac{-1}{\tau^{(i)}_0} )
+\imth(h^R_n-h^L_n)\Re (\frac{-1}{\tau^{(i)}_0} )
&=-\frac{\log (\la^{lr}_n)}{2\pi} +\text{ mod } \imth,
\nonumber\\
 (h^R_n+h^L_n-\frac{c}{12})\Im (\tau^{(i)}_1)
+\imth(h^R_n-h^L_n)\Re (\tau^{(i)}_1)
&=-\frac{\log (\la^{lurd}_n)}{\pi} +\text{ mod } 2\imth,
\nonumber\\
 (h^R_n+h^L_n-\frac{c}{12})\Im (\frac{-1}{\tau^{(i)}_1} )
+\imth(h^R_n-h^L_n)\Re (\frac{-1}{\tau^{(i)}_1} )
&=-\frac{\log (\la^{ldru}_n)}{\pi} +\text{ mod } 2\imth,
\end{align}
\end{widetext}
where $n=0,1,2,...$ and $\la^{ud}_n$, $\la^{lr}_n$, $\la^{lurd}_n$,
and $\la^{ldru}_n$ are the eigenvalues of $M^{ud}$, $M^{lr}$,
$M^{lurd}$, and $M^{ldru}$, respectively.  Also
\begin{align}
 \tau^{(i)}_1=
\frac{W^{(i)}_{21}+\imth W^{(i)}_{22} -W^{(i)}_{11}-\imth W^{(i)}_{12}}
{W^{(i)}_{11}+\imth W^{(i)}_{12}+W^{(i)}_{21}+\imth W^{(i)}_{22}}
=\frac{\tau^{(i)}_0-1}{\tau^{(i)}_0+1}
\end{align}
Since $h_0=0$, we see that the central charge $c$ and the scaling dimensions
$h^L_1,h^R_1,...$ can be calculated from the invariant fixed-point tensor
$T^{(i)}_\text{inv}$.

We like to stress that the invariant fixed-point tensor $T_\text{inv}$
can be calculated from \eq{Tinv} even away from the critical points.
This allows us to calculate the central charge $c$ and the scaling
dimensions $h^L_1,h^R_1,...$ from $T_\text{inv}$ away from the
critical point.  For phases with short-range correlations, the
resulting central charge $c$ will be zero.  However, the central
charge will be non-zero when correlation length diverges (or energy
gap vanishes).  Thus plotting central charge as a function of
temperature and other parameters is a good way to discover second
order phase transitions.  Also for phases with short-range
correlations, a finite number of $h^R_n,\ h^L_n$ are zero (which
represent the degenerate ground states) and rest of them are infinite.

We can also use the fixed-point tensors, to calculate the low energy
spectrum of a quantum system, such as our spin-1 system.\eq{spin1}
From the tensor-network representation of 1D quantum Hamiltonian (see
Fig. \ref{stepY}a), we see that the shape of the original tensor
network is described by a $W$-matrix
$ W=\bpm
1 & -1  \\
\del \tau & \del \tau \\
\epm $.
After $i$ steps of iteration, the $W$-matrix becomes $W^{(i)}=2^i W$.
From $W^{(i)}$, we find that $\tau_1^{(i)}$ is purely imaginary.  When
$\tau_1^{(i)}\propto \imth$, the eigenvalues of the matrix $-\log
(M^{lurd})$ will be proportional to the energy eigenvalues of our
spin-1 system,  as one can see from the expression of the partition
function \eq{Zch}.  From this result and using the fixed-point tensor,
we can calculate the energy spectrum of quantum system.

\subsection{Phases and phase transitions from the fixed-point tensors}
\label{phase}

In this section, we like to discuss how to define phases and phase
transition more carefully and in a very general setting.  For a system
that depends on some parameters, for example $\bt$ and $V$ for our
generalized loop-gas model, we can calculate the partition function
$Z(\bt,V)$ as a function of those parameters.  The phase and the phase
transitions are determined from the  partition function $Z(\bt,V)$ in
the following way:
The singularity in the partition function $Z(\bt,V)$ marks the
position of a phase transition. (This defines the phase transition.)
The regions separated by the phase
transition represents different phases.  In other words, if a system
$(\bt_1,V_1)$ and another system $(\bt_2,V_2)$ can be deformed into
each other without phase transition, then the two systems are in the
same phase.  If there is no way to deform system $(\bt_1,V_1)$ to
system $(\bt_2,V_2)$ without encountering at least one phase
transition, then the two systems are in different  phases.  We like to
point out that to avoid the complication of taking thermal dynamical
limit, here we only consider systems with some translation symmetry.
We also like to point out that the above definition is very general.
It includes symmetry breaking phases, topological phases, as well as
the phases with quantum order.\cite{Wqoslpub}

However, the above definition is still incomplete.  To define phases
and phase transitions, it is also very important to specify the
conditions that the Hamiltonian or Lagrangian must satisfy.  It is well
known that without specify the symmetry of Hamiltonian, the symmetry
breaking phases and symmetry breaking phase transitions cannot be
defined. After we specify the conditions on the Hamiltonian, the
deformations used in defining phases and phase transitions must
satisfy those conditions.

Since the conditions on the Hamiltonian or Lagrangian are part of the
definition of phases and phase transitions, it is not surprising that
the meaning of phases and phase transitions varies as we vary the
conditions on the systems.  Let us illustrate this point using the
generalized loop-gas model along the $\bt=0$ line.  In fact, weather
the fixed-point tensors $T^\text{LL}$ for $V<V_c$ and $T^{Z_2}$ for
$V>V_v$  in the generalized loop-gas model describe two different
phases or not depend on the conditions on the system.

Since we are using tensor network to describe our systems, the
conditions on the systems (or the Hamiltonian/Lagrangian) will be
conditions on the tensors.
If the only condition on the tensors is that the tensors $T_{rudl}$ must
be real, then $T^\text{LL}$ and
$T^{Z_2}$ describe the same phase. To show this, let us
consider a family of tensors labeled by $\th$
\begin{align*}
T^{\th}_{ruld} =  R^{-1}_{rr'}(\th) R^{-1}_{uu'}(\th)
T^{Z_2}_{r'u'l'd'} R_{l'l}(\th) R_{d'd}(\th)
\end{align*}
where
$ R(\th)=\bpm
\cos \th  & \sin \th \\
\sin \th  & -\cos \th
\epm $.
We see that the tensors $T^\th$ connect $T^{Z_2}=T^{\th=0}$ and
$T^\text{LL}=T^{\th=\pi/4}$ as we change $\th$ from $0$ to $\pi/4$.
Since $R(\th)$ is a field redefinition transformation, the partition
function for $T^\th$ does not depends on $\th$ and is hence a smooth
function of $\th$.  Therefore, $T^\text{LL}$ and $T^{Z_2}$ describe
the same phase.

On the other hand, if we specify that the tensors must satisfy the
closed loop condition \eq{loopT} (or \eq{Z2z}) and the $Z_2$ symmetry
condition \eq{Z2x}, then $T^\text{LL}$ and $T^{Z_2}$ describe two
different phases.  The continuous deformation $T^\th$ can no longer be
used.  This is because $T^\th$ does not satisfy \eq{loopT} if $0<\th <
\pi/4$.

Before ending this section, we like to point out that due to field
redefinition ambiguity in the TEFR calculation, some
time, a single phase may correspond to a group fixed-point tensors
that are related by field redefinitions.  To understand this point
more concretely, let us consider in more detail the $V>V_c$ phase
described by the fixed-point tensor $T^{Z_2}$.  In the decomposition
of rank-4 tensor $T$ into two rank-3 tensors $S_2$ and $S_4$:
\begin{equation*}
 T_{ruld}=\sum_a (S_2)_{ldk} (S_4)_{ruk}
\end{equation*}
we may have a ambiguity that $T$ can also be composed into
two different rank-3 tensors $S'_2$ and $S'_4$
\begin{equation*}
 T_{ruld}=\sum_a (S'_2)_{ldk} (S'_4)_{ruk}
\end{equation*}
where $S_2$ and $S'_2$, $S_4$ and $S'_4$ are related by an invertible
transformations
\begin{align*}
 (S_2)_{ldk}&=A_{kk'}(S'_2)_{ldk'}
&
 (S_4)_{ldk}&= (S'_4)_{ldk'} (A^{-1})_{k'k} .
\end{align*}
Here $A$ represents the field redefinition ambiguity in the TEFR
renormalization approach.

If the tensors $T$ and $S$ must satisfy certain conditions, such as
the closed loop condition \eq{loopT}, those conditions will limit the
ambiguity $A$.  For the $V>V_c$ phase, the fixed-point tensor is
$T^{Z_2}$. The closed loop condition \eq{loopT} limits the ambiguity
of the decomposition of $T^{Z_2}$ to a diagonal $A$ matrix.  The field
redefinition \eq{fred} does not change $T^{Z_2}$ when $A$ and $B$ in
\eq{fred} are diagonal matrices.  This way, we show that the
fixed-point tensor in the $V>V_c$ phase has no field redefinition
ambiguity.  Similarly, for the $V<V_c$ phase, the fixed-point tensor
$T^\text{LL}$ also has no field redefinition ambiguity, if we
implement the closed loop condition \eq{loopT}.  Those results are
confirmed by our numerical calculation.

\subsection{A detailed discussion of the renormalization flow of tensors}
\label{TEFflow}

Let us describe the TEFR flow in our spin-1 model in more detail which
allows us to see how an initial time-reversal and parity symmetric
tensor $T$ flows to the various fixed-point tensors.  Let us first
consider only the SVDTRG transformation described in Fig. \ref{CGsq}.

It is most convenient to describe the tensor in term of its singular
values $\la_s$ under the SVD decomposition (see \eq{ULaV}).  At first
a few steps of iterations, the singular values $\la_s$ are in general
not degenerate.  If the initial tensor $T$ is in the Haldane phase,
then after several iterations, the four largest singular values become
more and more degenerate.  If the initial tensor $T$ is in the TRI
phase, then after several iterations, the largest singular values
remain non degenerate.  We see that the Haldane phase and the TRI
phase can be distinguished from their different tensor flowing
behaviors under SVDTRG transformation.

Now let us consider the tensor flow under the TEFR transformation in
Fig. \ref{TEFsqCG}.  Again at first a few steps of iterations, the
singular values $\la_s$ are in general not degenerate.  Also the
tensors in first a few steps are quite different from CDL tensor.  As
a result, the scaling-SVD operation in Fig.
\ref{TEFsqCG}(b)$\to$(c) is a null operation which does not simplify
the tensors.  After several iterations, the tensors become closer and
closer to CDL tensors.  So after a certain step of iterations, the
scaling-SVD operation becomes effective which further
simplify the tensor.

If the initial tensor $T$ is in the TRI phase, even after the tensor
$T$ becomes CDL tensor, the singular values $\la_s$ are still not
degenerate.  This implies that the corner matrices $M_i$ in the CDL
tensors (see \eq{cdlts}) have non-degenerate eigenvalues.  The
entanglement filtering operation in Fig. \ref{TEFsqCG}(b)$\to$(c)
actually simplify the tensors by removing the smaller eigenvalues in
the corner matrices $M_i$.  So the scaling-SVD operation
reduces the corner matrices $M_i$ to a trivial one
$(M_i)_{ii'}=\del_{1,i}\del_{1,i'}$.  This way the scaling-SVD
operation reduces the initial tensor to the trivial
dimension-one tensor $T^\text{TRI}$.

If the initial tensor $T$ is in the Haldane phase, after the tensor
$T$ becomes CDL tensor after a few steps of TEFR iteration, the largest
four singular values $\la_s$ also become degenerate.  This implies
that the largest two eigenvalues of the corner matrices $M_i$ in the
CDL tensors are also degenerate.  In this case, the scaling-SVD
operation cannot reduce such a degenerate corner matrices to
a trivial one.  But it can still simplify the corner matrices to a
certain degree by removing other smaller eigenvalues.  This way, the
combinations of the SVD coarse graining in \ref{TEFsqCG}(d)$\to$(e)
and the entanglement filtering in Fig.  \ref{TEFsqCG}(b)$\to$(c)
reduce the initial tensor $T$ to a simple fixed-point tensor
$T^\text{Haldane}$.  Therefore the Haldane phase and the TRI phase can
be distinguished by their different fixed-point tensors under the TEFR
renormalization transformation.

We like to point out the emergence of the 4-fold degenerate singular
values requires the initial tensors to satisfy the time-reversal and
parity symmetries.  If we start with an initial tensor $T$ that has
time-reversal symmetry but not parity symmetry, our numerical results
show that the largest singular values do not become degenerate after
many iterations.  In this case the entanglement filtering operation
through the scaling-SVD always reduces the initial tensor $T$ to
$T^\text{TRI}$.  This implies that, without the parity symmetry, the
Haldane phase is the same as the trivial phase TRI.

We see that the emergence of 4-fold degenerate singular values is
closely related to the Haldane phase -- a topologically ordered phase.
In \Ref{LH0804}, an entanglement spectrum is introduced to identify
topological order.  We believe that the emergence of 4-fold degenerate
singular values is closely related to the degeneracy in the
entanglement spectrum.

We also like to point out that, only away from critical points, the
tensors flow to CDL tensor.  Near critical points, tensors are always
quite different from CDL tensors.  As a result, the entanglement
filtering operation in Fig.  \ref{TEFsqCG}(b)$\to$(c) is a null
operation which does not simplify the tensors.  So near a critical
point, the SVDTRG in Fig. \ref{CGsq} and the TEFR in Fig.
\ref{TEFsqCG} produce the same result.

\subsection{Projective realization of symmetries on fixed-point tensors}
\label{psg}

We like to point out that in the original tensor network of $T$ for the spin-1
system \eq{THi}, the time reversal and parity symmetry are given in \eq{timeT}
and \eq{ParityT}.  However, as discussed near the end of the Appendix
\ref{phase}, the TEFR transformation has a field-redefinition ambiguity.  As a
result, the time-reversal and parity transformations contain a
field-redefinition ambiguity, when act on the coarse grained tensor network.
Similarly, the fixed-point tensor also contain such field-redefinition
ambiguity.  If we choose a basis that the fixed-point tensor in the Haldane
phase is given by $T^\text{Haldane}=T( \si^y, \si^y, \si^y, \si^y)$, the time
reversal and parity transformations that act on the fixed-point tensor are
given by
\begin{widetext}
\begin{align}
\text{Parity: } & T \to T_p, \text{ where }  (T_p)_{ruld}=
\sum_{r'u'l'd'}
W_{rr'}
W_{uu'}
W_{ll'}
W_{dd'}
T_{u'r'd'l'}.
\label{Sparity}\\
\text{Time reversal: } & T \to T_t, \text{ where }
(T_t)_{ruld}=T_{dlur}.\label{Stime}
\end{align}
\end{widetext}
Here $W$ is a field-redefinition transformation
\begin{equation}
 W=\bpm
1 & 0 & 0 & 0 \\
0 & 0 & 1 & 0 \\
0 & 1 & 0 & 0 \\
0 & 0 & 0 & 1 \\
\epm .\label{redfine}
\end{equation}
The time-reversal and parity transformations on the coarse grained
tensor network and those on the initial tensor network (see \eq{timeT}
and \eq{ParityT}) differ by a field-redefinition transformation.  The
fixed-point tensor $T^\text{Haldane}$ is invariant under the above
projective symmetry group (PSG) transformations.\cite{Wqoslpub} We
also numerically confirm that the fixed-point tensor
$T^\text{Haldane}$ is stable if it is perturbed by a tensor $\del T$
that is invariant under the same PSG defined through \eq{Sparity} and
\eq{Stime}.  We note that the PSG is used to define quantum order (a
generalization of topological order) in \Ref{Wqoslpub}.
The tensor network renormalization naturally allows
the generalization from topological order to quantum order.

We like to remark that the field-redefinition transformation should be derived
by keeping track of the parity and time reversal symmetry at each TEFR flow
step. As a result, the field-redefinition transformation $W$ should
satisfy \eq{Sparity} for the fixed point tensor $T^\text{Haldane}=T(
\si^y, \si^y, \si^y, \si^y)$.
However,
not all the $W$ that satisfy
(\ref{Sparity}) for the fixed point tensor  $T^\text{Haldane}=T(
\si^y, \si^y, \si^y, \si^y)$ correspond to the parity symmetry for
the original tensor \eq{ParityT}. For example:
\begin{equation}
 W^\prime=\bpm
0 & 0 & 0 & 1 \\
0 & -1& 0 & 0 \\
0 & 0 & -1 & 0 \\
1 & 0 & 0 & 0 \\
\epm .\label{redfine1}
\end{equation}
also satisfies Eq. \eq{Sparity}, but this is not the correct
field-redefinition transformation for parity symmetry of
$T^\text{Haldane}=T( \si^y, \si^y, \si^y, \si^y)$.
In particular, an
perturbations with the wrong parity symmetry:
\begin{align}
\del T_{ruld}= \sum_{r'u'l'd'} W_{rr'}^\prime W_{uu'}^\prime
W_{ll'}^\prime W_{dd'}^\prime \delta T_{u'r'd'l'}\label{Sparity1}
\end{align}
and time reversal symmetry \eq{Stime} can destabilize the fixed-point tensor
$T^\text{Haldane}=T(
\si^y, \si^y, \si^y, \si^y)$.

For the fixed point tensor of Haldane phase, the field-redefinition
transformations that correspond to parity and time reversal symmetries are also
base dependent. If way choose $T^\text{Haldane}=T( \si^0, \si^0, \si^0,
\si^0)$, then the correct field-redefinition transformation corresponding to
the parity symmetry should be $W^\prime$ not $W$. We can also choose other
bases that $T^\text{Haldane}$ is invariant in a trivial way under parity
transformation  but invariant up to a field-redefinition under time reversal
transformation.


\end{document}